\documentclass[twocolumn,english,aps,prx,floatfix,amssymb,superscriptaddress,longbibliography]{revtex4}
\usepackage[latin9]{inputenc}
\usepackage{verbatim}
\usepackage{float}
\usepackage{amsmath}
\usepackage{amssymb}
\usepackage{graphicx}
\usepackage{color}
\usepackage[bookmarks=false,linkcolor=blue,urlcolor=blue,colorlinks,citecolor=blue]{hyperref}
\makeatletter

\newcommand{\be}{\begin{equation}}
\newcommand{\ee}{\end{equation}}
\newcommand{\bea}{\begin{eqnarray}}
\newcommand{\eea}{\end{eqnarray}}

\@ifundefined{textcolor}{}
{%
 \definecolor{BLACK}{gray}{0}
 \definecolor{WHITE}{gray}{1}
 \definecolor{RED}{rgb}{1,0,0}
 \definecolor{GREEN}{rgb}{0,1,0}
 \definecolor{BLUE}{rgb}{0,0,1}
 \definecolor{CYAN}{cmyk}{1,0,0,0}
 \definecolor{MAGENTA}{cmyk}{0,1,0,0}
 \definecolor{YELLOW}{cmyk}{0,0,1,0}
}

\usepackage[caption=false]{subfig}

\@ifundefined{showcaptionsetup}{}{%
 \PassOptionsToPackage{caption=false}{subfig}}
\usepackage{subfig}
\makeatother

\usepackage{babel}
\begin{document}

\title{From one-dimensional charge conserving superconductors to the gapless
Haldane phase}
\author{Anna Keselman}
\affiliation{Station Q, Microsoft Research, Santa Barbara, California 93106-6105, USA}
\affiliation{Department of Condensed Matter Physics, Weizmann Institute of Science, Rehovot, Israel 7610001}
\author{Erez Berg}
\affiliation{James Frank Institute, The University of Chicago, Chicago, IL 60637, USA}
\affiliation{Department of Condensed Matter Physics, Weizmann Institute of Science, Rehovot, Israel 7610001}
\author{Patrick Azaria}
\affiliation{Laboratoire de Physique Thorique des Liquides, Universit Pierre et Marie Curie, 4 Place Jussieu, 75252 Paris, France}
\affiliation{Department of Condensed Matter Physics, Weizmann Institute of Science, Rehovot, Israel 7610001}

\begin{abstract}
We develop a framework to analyze one-dimensional topological superconductors with charge conservation. 
In particular, we consider models with $N$ flavors of fermions and $(\mathbb{Z}_2)^N$ symmetry, 
associated with the conservation of the fermionic parity of each flavor.
For a single flavor, we recover the result that a distinct topological phase with exponentially localized zero modes does not exist
due to absence of a gap to single particles in the bulk. 
For $N>1$, however, we show that the ends of the system can host low-energy, exponentially-localized modes. 
The analysis can readily be generalized to systems in other symmetry classes.
To illustrate these ideas, we focus on lattice models with $SO\left(N\right)$ symmetric interactions, 
and study the phase transition between the trivial and the topological gapless phases 
using bosonization and a weak-coupling renormalization group analysis.
As a concrete example, we study in detail the case of $N=3$. 
We show that in this case, the topologically non-trivial superconducting phase corresponds to a gapless analogue of the Haldane phase in spin-1 chains.
In this phase, although the bulk is {\it gapless} to single particle excitations,
the ends host spin-$1/2$ degrees of freedom which are exponentially
localized and protected by the spin gap in the bulk. 
We obtain the full phase diagram of the model numerically, using density matrix renormalization group calculations. 
Within this model, we identify the self-dual line studied by Andrei and Destri
[Nucl. Phys. B, \textbf{231}(3), 445-480 (1984)], 
as a first-order transition line between the
gapless Haldane phase and a trivial gapless phase. This allows us to
identify the propagating spin-$1/2$ kinks in the Andrei-Destri model
as the topological end-modes present at the domain walls between the
two phases.
\end{abstract}
\maketitle

\section{Introduction}

One-dimensional topological superconductors have been in the focus
of both experimental and theoretical study in condensed matter physics
in the past decade, due to the unique excitations, Majorana bound
states, they host at their ends \cite{Alicea2012,Beenakker2013,leijnse2012introduction}. 
The topological protection of these
modes relies on the bulk of the system being gapped. This is the case
if the one-dimensional system is proximity coupled to a bulk superconductor.
However, in a purely one-dimensional system, in which superconductivity
arises from intrinsic attractive pairing interactions, the bulk of
the system remains gapless due to large quantum phase fluctuations.

In spin-polarized systems with intrinsic pairing interactions, Majorana end-modes are, generically, no longer protected \cite{Fidkowski2011,Sau2011} 
since the bulk is gapless to single particle excitations \cite{Ruhman2015,kane2017pairing}.
In this work, we discuss situations where well-defined, exponentially localized
end-modes can survive in a purely one-dimensional system in the presence of additional symmetries. 
Several examples of such phases are known \cite{Starykh2000, Starykh2007, Ruhman2012, Zoller2013, Keselman2015, Diehl2015, chen2017flux, kainaris2017interaction, kainaris2017transmission, Scaffidi2017}.
However, a general framework that relates these phases to their non-charge conserving (mean field) counter parts has not been given.

Here, we develop a general approach to treat one-dimensional
systems with multiple fermion flavors and intrinsic, charge-conserving, attractive
interactions. As a test case, we study a family of models with $(\mathbb{Z}_2)^N$ symmetry, 
associated with the conservation of the fermionic parity of each flavor.
We show how the result for the spinless case ($N=1$) can be recovered using this approach.
We then address the case of $N>1$ flavors, and show that in this case, 
the system can host low-energy exponentially localized end-modes,
reminiscent of the Majorana zero modes, despite the gaplessness of the bulk.

We discuss the low-energy structure of the gapless topological phases for arbitrary $N$ 
and emphasize the connection between the nature of the low-energy modes in the bulk and the protection of the end-modes.
We show that for odd $N$, the bulk hosts low-energy composite fermionic excitations, with momentum $N k_F$. For even $N$ the bulk hosts only bosonic excitations.

Further, we present a more detailed analysis of lattice models with $SO\left(N\right)$ symmetric interactions. 
In this case, the protection of the end modes can be understood as arising from the presence of a spin-gap in the bulk. 
For $N=2$, we recover the phase studied previously
in  \cite{Keselman2015}, where it was shown that the protection
is in fact robust to breaking of $SO\left(2\right)$ symmetry, as
long as time-reversal symmetry is present. For $N=3$, we show that
the topological superconducting phase is closely related to the Haldane
phase in spin-1 chains. In this phase, although the bulk is gapless
to single particle excitations, the ends host spin-1/2 degrees of
freedom which are exponentially localized and protected by a spin
gap in the bulk \cite{jiang2017symmetry}. 

This paper is organized as follows. We start by presenting the model
under consideration in Sec. \ref{sec:StabilityTSC}, discussing the
connection to fully-gapped topological superconductors 
and addressing the stability of the topological phase to quantum
phase fluctuations based on general heuristic arguments. We explain the connection between the nature of the bulk low-energy modes and the topological protection of the end-modes.
In Sec. \ref{sec:lowenergy}, we analyze the low-energy physics of the gapless topological phase in greater detail, distinguishing between the cases of even and odd $N$.
In Sec. \ref{sec:SON}, we focus on $SO\left(N\right)$ symmetric models. 
We present a field theory analysis backing up the heuristic arguments presented previously.
To this end, we discuss a slightly generalized model with on-site
interactions, which hosts both a trivial and a topological gapless
phase. We discuss a duality transformation relating these phases,
and the phase diagram expected from weak coupling RG. 
In Sec. \ref{sec:GaplessHaldane} we study in detail the lattice model for $N=3$, with both on-site
and nearest-neighbors $SO\left(3\right)$ symmetric pairing interactions, and map out its phase diagram using the density matrix renormalization group (DMRG) \cite{White1992PRL,White1992PRB}. The topological superconducting phase in this model is identified as a gapless analogue of the symmetry-protected Haldane phase of $S=1$ spin chains. The conclusions are summarized in Sec.~\ref{sec:conclusions}.
\section{Stability of topological superconductors in 1D charge conserving
systems \label{sec:StabilityTSC}}

We start by describing a general heuristic argument to analyze the
stability of topological phases in one dimensional charge-conserving
superconductors and their protected edge modes.

\subsection{Spinless Wire}
\label{sec:spinless}

Consider a one-dimensional system of spinless fermions with a general short-range Hamiltonian that conserves the total charge. If the density of particles is incommensurate with the lattice, the system is generally gapless \cite{OshikawaAffleck}. Our goal is to map the possible distinct phases of the system, and to understand their low-energy properties; in particular, we ask about the nature of low-energy gapless modes in the bulk, and whether there are any well-defined zero modes bound to the edges. 

The problem of a single ``flavor'' of fermions with no symmetries other than charge conservation has been analyzed by various authors~\cite{Fidkowski2011,Sau2011,Ruhman2012,ruhman2017topological}; in this case, two distinct gapless phases are possible, one with a gap to single fermions in the bulk (but no gap to pairs of fermions) and the other with no gap to single fermions. Neither of these phases supports exponentially localized edge modes. Below, we derive this result using a general, heuristic argument; the argument is then easy to generalize to systems with additional symmetries.

For concreteness, it is useful to consider the following simple Hamiltonian:
\begin{equation}
H=\sum_{i}\left(-tc_{i}^{\dagger}c_{i+1}+h.c.-\mu c_{i}^{\dagger}c_{i}\right)+\sum_{i,j}V_{ij}\Pi_{i}^{\dagger}\Pi_{j},\label{eq:H_single}
\end{equation}
where $\Pi_{i}^{\dagger}=c_{i}^{\dagger}c_{i+1}^{\dagger}$, and $V_{ij}<0$
denote the strength of attractive interactions. We will assume that
the interactions are short-ranged, but not necessarily nearest neighbor.
Our considerations will be much more general, and apply to any one-dimensional Hamiltonian of spinless fermions (including further neighbor hopping, other forms of interactions, etc.)

We formulate the problem as a path integral and decouple the interaction
term via a Hubbard-Stratonovich transformation
\begin{equation}
Z=\int\prod_{i}Dc_{i}D\bar{c}_{i}D\Delta_{i}e^{-\int d\tau L},
\end{equation}
where $c_{i},\bar{c}_{i}$ are Grassmann variables and the Lagrangian
is given by
\begin{multline}
L=-\sum_{i}\bar{c}_{i}\left(\partial_{\tau}-\mu\right)c_{i}\\
+\sum_{i}\left(-t\bar{c}_{i}c_{i+1}+\Delta_{i}\bar{c}_{i}\bar{c}_{i+1}+h.c.\right)\\
+\sum_{i,j}\left(V^{-1}\right)_{ij}\Delta_{i}\Delta_{j}^{*}.\label{eq:L}
\end{multline}
We assume fluctuations in the amplitude of $\Delta_{i}$ are small and denote
$\Delta_{i}=\left|\Delta\right|e^{2i\theta_{i}}$. The last term in
(\ref{eq:L}) gives rise to a finite phase stiffness of $\theta$,
allowing us to assume that it changes slowly between neighboring
sites. Hereafter, we assume the phase varies slowly, but omit this
term for simplicity.

If the attractive interactions are sufficiently strong and long-ranged, then the phase $\theta_i(\tau)$ fluctuates slowly in space and time. 
Approaching the problem from this quasi-ordered limit, we introduce neutral fermions $f_{i}=c_{i}e^{-i \theta_i}$.
(Note that under a $U\left(1\right)$ gauge transformation $c_{i}\to c_{i}e^{i\alpha}$,
$\theta_{i}\to\theta_{i}+\alpha$ leaving the $f_{i}$ fermions unchanged.) Since the phase varies little on the scale of a lattice spacing, 
the pairing and hopping terms can be approximated as $\Delta_{i}\bar{c}_{i}\bar{c}_{i+1}=\Delta\bar{f}_{i}\bar{f}_{i+1}e^{i\left(\theta_{i}-\theta_{i+1}\right)}\approx\Delta\bar{f}_{i} \bar{f}_{i+1}$ and $t \bar{c}_{i} c_{i+1} =   t e^{i(\theta_{i+1} - \theta_i)} \bar{f}_{i} f_{i+1} \approx t \bar{f}_{i} f_{i+1}$. 

Special care needs to be taken when considering the boundary conditions of the $f_i$ fermions on a closed ring. Phase configurations where the phase $\theta_i$ winds by $n_\theta \pi$, where $n_\theta \in \mathbb{Z}$ (i.e., $\theta_{N_x} \approx \theta_1 + n_\theta \pi$, where $N_x$ is the number of sites) should be accounted for. To avoid a discontinuity in the parameters of the Hamiltonian across the bond from site $N_x$ to $1$, we can encode the winding number in the boundary conditions of the fermions: $f_{N_x+1} = (-1)^{n_\theta} f_1$. 

After the transformation to the $f_i$ fermions, and assuming a nearly-static phase configuration, the Lagrangian is written as:
\begin{eqnarray}
L&=&\sum_{i}\left[\bar{f}_{i}\left(\partial_{\tau}-\mu\right)f_{i}+in_{i}\partial_{\tau}\theta_{i}\right] \nonumber \\
&+&\sum_{i}\left(- t  \bar{f}_{i} f_{i+1}+\Delta\bar{f}_{i}\bar{f}_{i+1}+h.c.\right),
\label{eq:Lf}
\end{eqnarray}
where $n_{i}=f_{i}^{\dagger}f_{i}=c_{i}^{\dagger}c_{i}$ is the occupation
of site $i$, and the boundary conditions of the fermions are periodic (anti-periodic) if the winding number of the phase is even (odd), respectively. Hence, the problem of finding the phase structure of the Hamiltonian \eqref{eq:H_single} has been mapped to the problem of classifying the possible phases of the $f$-fermions, that obey a mean-field like static Hamiltonian with a dynamically determined boundary condition.

For $-2t<\mu<2t$, the $f$ fermions realize a non-trivial, class-D
\cite{AltlandZirnbauer1997} topological superconductor, i.e. a Kitaev
chain. However, we argue below that the topological properties of the
phase (and in particular, its protected zero modes) are lost due the coupling between the $f$ fermions
and the phase winding of the superconducting order parameter, $n_\theta$. 
This coupling allows for low-energy single particle excitations in the bulk of the system, unlike the mean-field case, where single fermions are gapped in the bulk. Such excitations can couple between the Majorana modes at the ends of the chain, removing the exponential ground state degeneracy.

To see that the bulk is gapless to single particle excitations consider a system with periodic boundary conditions and ${\cal N}$ fermions. Recall that a Kitaev chain on a closed ring has a unique ground state with a well defined fermion parity \cite{Kitaev2001}. Adding a single fermion excites the system to energy $\Delta$, the magnitude of the superconducting gap. However, in our case, this gap can be avoided by introducing a $\pi$ phase winding in $\theta$, changing the boundary conditions of the $f$ fermions from periodic to anti-periodic or vice versa. The Kitaev chain's ground state has an opposite fermion parity with periodic and anti-periodic boundary conditions~\cite{Kitaev2001}.
Therefore, the energy cost of adding a single fermion is only due to the extra phase winding (which costs an energy proportional to the inverse of the system size; this is nothing but the charging energy). 
Due to the presence of low energy single fermion excitations in the bulk, the end modes are no longer exponentially localized; they can leak into the bulk and the localization becomes power-law, with a power dictated by the Luttinger parameter in the wire~\cite{Sau2011,Fidkowski2011}.

If the $f$ fermions in (\ref{eq:Lf})
realize a trivial superconducting phase,
the superconducting gap $\Delta$ cannot be avoided when an extra fermion is added to the system, since changing the boundary conditions no longer changes the fermion parity of the ground state. Hence, the bulk is gapped to single particle excitations. However, in this case there are no low-energy end modes. 
This establishes that the single-flavor chain ($N=1$) supports two distinct phases, as has been discussed in Refs. \cite{ruhman2017topological,kane2017pairing}, neither of which supports exponentially localized end-modes.

The same conclusion can be reached from a bosonization analysis.
The system in this case maps to a single-flavor Luttinger liquid which has no gap to single fermions.
However, the argument above is more general: it implies that, as a matter of principle \footnote{This conclusion does not rely on any particular field theoretic formulation, it follows directly from the action \eqref{eq:Lf}.}, a topological
phase with a single Majorana end mode cannot be stable if charge is
conserved, since the bulk must be gapless to single fermion excitations. 

\subsection{Wire with $N$ Flavors\label{subsec:N_flav}}

Let us now study a less trivial situation where
several flavors of fermions are present. We shall consider a model with $N$ flavors, $c_{a=1,\dots,N}$, each with a Hamiltonian of the form of
Eq.~\eqref{eq:H_single}. The different flavors are coupled by a pair hopping term. The total Hamiltonian is given by
\begin{equation}
H= \sum_{a=1}^N H_a + H_{\rm int},
\end{equation} 
where
\begin{eqnarray}
H_a&=&\sum_{i}\left(-tc_{i,a}^{\dagger}c^{}_{i+1,a}+h.c.-\mu c_{i,a}^{\dagger}c^{}_{i,a}\right) \nonumber \\ &+& \sum_{i,j}  V^a_{ij} \;  \Pi_{i,a}^{\dagger} \Pi^{}_{j, a},  \nonumber \\ 
H_{\rm int} &=&  \sum_{i,j,a,b}V_{ij}^{ab}\; \Pi_{i,a}^{\dagger} \Pi^{}_{j ,b},
\label{HNflavors}
\end{eqnarray}
where  $\Pi_{i, a}^{\dagger}=c_{i, a}^{\dagger}c_{i+1, a}^{\dagger}$ is a p-wave pair creation operator
for a single flavor and $V^a_{ij} < 0$ denotes the attractive interaction within each flavor channel.
The pair hopping couplings, $V_{ij}^{ab}=V_{ij}^{ba}$,  are assumed to be short ranged.  We shall also  restrict ourselves to attractive interactions  $V_{ij}^{ab} <0$.

The Hamiltonian (\ref{HNflavors}) has a global $U(1)$ symmetry associated with total charge conservation and a $(\mathbb{Z}_2)^N$ symmetry associated with the conservation of the fermionic parity of each flavor. 
Introducing  the fermionic parity operator associated with a flavor $a$, 
 \begin{equation}
 P_a=(-1)^{n_a}, \; n_a= \sum_i c_{i,a}^{\dagger}c^{}_{i,a}, 
 \label{eq:Fparityflavor}
 \end{equation}
we see indeed that  $[P_a, H]=0,\; a=(1,...,N)$.
The above symmetry is of utmost  importance in regards to the topological properties of (\ref{HNflavors}).  Indeed, as shown in  \cite{Sau2011}, if not for the above $(\mathbb{Z}_2)^N$ symmetry, the topological character of the $N$ Kitaev chains would not survive when charge conservation is imposed. As we shall argue   below, the  $(\mathbb{Z}_2)^N$ symmetry (\ref{eq:Fparityflavor}) protects the end modes associated with each chain from leaking into the bulk and allows, even in a charge conserving system, for symmetry protected topological (SPT) phases for all $N>1$. In all what follows, we shall assume that the  $(\mathbb{Z}_2)^N$ symmetry (\ref{eq:Fparityflavor}) is present and keep the discussion as general as possible, allowing $ V^a_{ij} < 0$ and $V_{ij}^{ab}< 0$ to be arbitrary.

\subsubsection{Bulk spectrum}
  
Following the same procedure as in Sec.~\ref{sec:spinless}, we introduce Hubbard-Stratonovich
fields $\Delta_{i,a} = \left| \Delta_{i, a} \right| e^{2i\theta_{i, a}}$ and
neutral fermions $f_{i, a}=c_{i, a}e^{-i\theta_{i, a}}$. We neglect the amplitude fluctuations of the superconducting order parameters, replacing $|\Delta_{i,a}|$ by $\Delta_a>0$.  
The Lagrangian takes the form
\begin{equation}
L= \sum_{a=1}^{N} L_a + L_{\rm int},
\label{eq:LN}
\end{equation}
where $L_a$ is the Lagrangian associated with a single Kitaev chain of flavor $a$
\begin{eqnarray}
L_a&=&\sum_{i}\left[\bar{f}_{i, a}\left(\partial_{\tau}-\mu\right)f_{i, a}+in_{i, a}\partial_{\tau}\theta_{i, a}\right]  \\
&+& \sum_{i}\left(-t \bar{f}_{i, a}f_{i+1, a}+ \Delta_a  \;  \bar{f}_{i, a}\bar{f}_{i+1, a}+h.c.\right),\nonumber\label{eq:Lf-N}
\end{eqnarray}
and 
\begin{equation}
L_{\rm int}=  \sum_{i, j}\sum_{ a\neq b} J_{ij}^{ab}\ \cos{2(\theta_{i,a}- \theta_{j, b})},
\label{eq:locking}
\end{equation}
where $J_{ij}^{ab} =2\Delta_a \Delta_b V_{ij}^{ab}$, are Josephson couplings which tend to lock the phases $\theta_{i,a}$ together. As in the single-flavor case, on a ring with $N_x$ sites, the fermions satisfy the boundary conditions: $f_{N_x+1,a} = (-1)^{n_{\theta,a}} f_{1,a} $, where $n_{\theta,a}$ is the winding number of the superconducting phase of flavor $a$: $\theta_{N_x,a} \approx \theta_{1,a} + n_{\theta,a} \pi$ [see discussion above Eq.~\eqref{eq:Lf}].

Assuming that  the phases $\theta_{i,a}$ are slowly varying, one may integrate out the $f$ fermions. In the low-energy limit  the Lagrangian (\ref{eq:LN}) reduces  to \cite{Sau2011}
\begin{equation}
L= \int dx \,  [   \sum_{a=1}^N \frac{K_a}{2 \pi} (\partial_{\mu} \theta_a)^2 +  \sum_{a\neq b=1}^N G^{ab} \cos{2(\theta_{a}(x)- \theta_{ b}(x))}],
\label{eq:LuttN}
\end{equation}
where  $\partial_{\mu} \equiv (\partial_{x}, \partial_{ \tau})$ (we set the velocity scale $v\equiv 1$),  $K_a$ are Luttinger parameters associated with the chains and $G^{ab} < 0$ are effective Josephson couplings that lock the phases to each other such that $\theta_{a}(x)- \theta_{b}(x) =(p_a - p_b) \pi$, $p_a \in \mathbb{Z}$. 
 
Consider now adding a single fermion of a given flavor to a system on a ring. As in the spinless case analyzed above, the energy cost of adding a single fermion of flavor $a$ can be avoided by introducing a phase winding in the field $\theta_{ a}$. However, as the phases $\theta_{1,..,N}$ are now locked to each other by the Josephson couplings (\ref{eq:locking}), such a phase winding costs a finite energy  $\Delta_{\sigma}$. We conclude, therefore, that for $N>1$ there is a gap to adding a fermion of a single flavor.
The same reasoning can be applied to excitations involving adding or removing $M$ fermions of different flavors, and leads to the conclusion that they are all gapped unless $M=N$. 

An excitation involving adding (or removing) a single fermion of \emph{each} flavor, however, is gapless (i.e. costs energy proportional to the inverse of the system size). To see this, consider introducing a phase winding \emph{simultaneously} in all the fields $\theta_{ a}$, thus avoiding the spin gap $\Delta_{\sigma}$.
Doing so switches the boundary conditions of fermions of all the flavors $f_{ a}$, while at the same time the fermionic parities of all $N$ flavors are changed, $P_a \rightarrow - P_a$ ($a=1,..., N$), thus avoiding the superconducting gap $\Delta_a$ in each chain. The only energy cost then is due to the phase twist which scales as $1/L$.

We therefore conclude that there are two types of gapless excitations in the system. Gapless excitations of the first type involve adding (or removing) pairs of fermions of the \emph{same} flavor, without changing the fermionic parity of any flavor (e.g. by acting with the p-wave creation operator $\Pi^{\dagger}_{i,a}$). Excitations of the second type are composite operators involving $N$ fermions of \emph{different} flavors that change the parity of all flavors.
In this respect the composite operators
\begin{equation}
{\tilde c}^{\dagger}_{i,Q}\sim c_{i, a_1}^{\dagger}... c_{i, a_M}^{\dagger} c_{i, b_1}^{}... c_{i, b_{N-M}}^{},
\label{eq:compositek}
\end{equation}
which have charge $Q=2M-N$, have a finite overlap with the gapless modes of the system. These composite excitations are fermionic for $N$ odd and bosonic for $N$ even. As a consequence, when $N$ is even the low-energy excitations are always bosonic whereas, when $N$ is odd, there exist both fermionic and bosonic excitations: while the fermions change the fermionic parity of all flavors, the bosons do not.
 
\subsubsection{End modes}\label{sec:endmodes}
 
Let us now investigate the fate of the end modes in a system with $N$ flavors, in light of preceding discussion.
If we neglect the fluctuations of the phase in the action of the $f_a$ fermions [Eq.~(\ref{eq:Lf-N})] then for $-2t<\mu<2t$, each flavor hosts a pair of zero-energy Majorana modes, $\gamma_{L, a}$ and $\gamma_{R, a}$, localized at the two ends of an open chain. We shall argue that the low-energy excitations in the bulk cannot couple the zero modes of the two edges; therefore, a system with open boundary conditions has exponentially localized edge modes, even when the phase fluctuations are taken into account. 

To see this, recall that the operators that create gapless bulk excitations either preserve the fermionic parity of each flavor separately, or change the fermionic parity of \emph{all} flavors. In the first case, their action on the low-energy Hilbert space associated with each edge has to be proportional to the identity operator,  since any non-trivial product of $\gamma_{L(R),a}$'s that act on the left (right) edge changes the parity of at least one flavor.
Operators of the form (\ref{eq:compositek}), that change the fermionic parity of all the flavors, must act on the low-energy Hilbert space of the left (right) edge as $\Pi_{a=1}^N  \gamma_{L(R), a}$. Hence, the coupling between the two ends is necessarily proportional to total fermionic parity:
\be
P= \Pi_{a=1}^N P_a = \Pi_{a=1}^N (-i \gamma_{L,a} \gamma_{R,a}).
\label{eq:Fparity}
\ee
Consequently, this coupling does not lift the topological degeneracy within a given parity sector.

Finally, we discuss the topological degeneracy associated with the edge modes. 
As usual, we label the states associated with the edges by the occupation
number of the complex (Dirac) fermions $d_a= (\gamma_{L, a}+ i\gamma_{R, a})/2$. These states can be separated
into two sets of $2^{N-1}$ states of even and odd total fermionic parity. In a system with overall conservation of the number of fermions, states with opposite fermionic parity must also have different charges; therefore,
we conclude 
that the only effect of charge conservation on the low-energy part of the spectrum is to lift the degeneracy between 
the even and odd parity states by an amount proportional to the charging energy of the system, which 
scales like $1/L$. The low-energy manifold in each charge sector contains $2^{N-1}$ states whose energy separation is exponentially small in the size of the system. 

The Majorana zero modes at each edge satisfy the Clifford algebra, $\{\gamma_{L,a}, \gamma_{L,b}\}= 2 \delta_{a,b}$ and $\{\gamma_{R,a}, \gamma_{R,b}\}= 2 \delta_{a,b}$.  
Therefore, the low-energy Hilbert space of each edge can be described as a $SO(N)$ spinor, $| \alpha_L \rangle $ and $| \alpha_R \rangle$. 
When $N$ is odd, there is a single irreducible spinor representation $\alpha_{L(R)}$ of dimension $2^{(N-1)/2}$
of the $SO(N)$ group generated by
\be
S_{L(R), ab}= \frac{1}{4i} [\gamma_{L(R),a}, \gamma_{L(R),b}],
\label{spinor}
\ee
and, in a given parity sector, the $2^{N-1}$ topological degeneracy is exhausted by the tensor product states $|\alpha_{L}\rangle \otimes |\alpha_{R}\rangle$. When $N$ is even the situation is more subtle. Now, there are 
two irreducible spinor representations of \eqref{spinor}  of dimension $2^{N/2-1}$. The spinors that belong to the two representations,
$|\alpha_{L(R), \pm}\rangle$,  are eigenvectors of the edge fermionic parity operators $P_{L(R)}= (-i)^{N/2} \Pi_{a=1}^N  \gamma_{L(R),a}$: $P_{L(R)} |\alpha_{L(R), \pm}\rangle = \pm |\alpha_{L(R), \pm}\rangle$. For $N$ even,
the fermionic parity operator \eqref{eq:Fparity} can be written as  $P=(-1)^{N/2} P_L P_R$ with $[P_L, P_R]=0$.
Therefore, the low energy subspace in a given parity sector is spanned by the tensor products $|\alpha_{L, \pm}\rangle \otimes |\alpha_{R, \pm} \rangle$ for $N/2$ even and $|\alpha_{L, \pm}\rangle \otimes |\alpha_{R, \mp}\rangle$ for $N/2$ odd.

Topological phases of charge-conserving 1d superconductors in other symmetry classes can be understood in a similar way to the arguments laid out above. For example, in class DIII (time reversal with $\mathcal{T}^2=-1$) \cite{AltlandZirnbauer1997}, the $f$ fermions carry spin $1/2$. The $f$-fermions then form one of the two distinct phases of mean-field Hamiltonians in class DIII. In either phase, twisting the boundary conditions does change the fermion parity of the ground state; therefore, there is a gap to single fermions in the bulk. However, in the topological phase each edge supports a topologically protected Kramers' pair, where the two states have an opposite local fermion parity \cite{Keselman2015}.

\section{Low-energy description of the topological phase \label{sec:lowenergy}}

As seen in the above discussion, the stability of the topological phase is ensured by both
the $(\mathbb{Z}_2)^N$ symmetry and the particular nature of the gapless
modes of the system: the fact that the operators that create these modes are either even under the fermionic parities of all the flavors or odd under all of them, is essential in protecting 
the zero modes at the ends. We now turn to the more familiar description in terms of the effective low-energy bosonized theory, making contact with the considerations above.  

As before, we assume that in the low-energy limit the phases associated with different flavors are locked
together. Denoting the collective phase by $\Theta(x, \tau)$ the effective low-energy
 Lagrangian of the system is that  of a generalized Luttinger liquid \cite{AzariaLuttinger}
\begin{equation}
L_{Lutt}= \frac{K}{2 \pi} \int dx \;   (\partial_{\mu} \Theta)^2,
\label{eq:Lutt}
\end{equation}
where $K$ is the Luttinger parameter.  {As the $\Theta$} field in (\ref{eq:Lutt}) is conjugate
to the total density of particles, $n(x)= \sum_{a=1}^N n_a(x)$,
\begin{equation}
[n(x), \Theta(y)]= i  \delta(x-y),
\label{eq: conjugate}
\end{equation}
we define the dual field $\Phi(x)$, as $n(x)  \equiv \partial_x \Phi(x)$, with
$[\Phi(x) ,  \Theta(y)]= i  Y(y-x)$, $Y(u)$ being the Heaviside step function. 
The $\Phi(x)$  and $\Theta(x)$ fields are related to the flavor bosonic fields,
$\phi_a(x)$ and  $\theta_a(x)$,  associated with the Luttinger liquids
describing  each flavor (\ref{eq:LuttN}) by the canonical transformation
\begin{equation}
\Theta=\frac{1}{N} \sum_{a=1}^N \theta_a, \; \;  \Phi = \sum_{a=1}^N \phi_a.
\label{eq: Thetatheta}
\end{equation}
These fields are related to the total charge $Q$ and current $J$,
\begin{equation}
J =\frac{N}{\pi} \int_0^L \partial_x \Theta, \; \;  Q=\int_0^L \partial_x \Phi.
\label{eq: QJ}
\end{equation}
We therefore find that  the total current of the system  is quantized in units of $N$. 
As we show below, this is a consequence of both the presence of a flavor gap
and of the topological nature of the phase.

The gapless excitations of the generalized Luttinger liquid (\ref{eq:Lutt}) can be expressed in terms of the vertex operators
\begin{equation}
V_{n,m}(x) \propto e^{i[n \Theta(x) + m \pi  \Phi(x)]},
\label{eq:vertex}
\end{equation}
which have the scaling dimension $\Delta_{nm}= (n^2/K + m^2 K)/4$. They 
 carry   charge and current
 \be
 Q=n,\; J=mN,\; (n,m)\in\mathbb{Z},
 \label{QJ}
 \ee
which owing to the relation between current and momentum have momentum $P=J k_F=mNk_F$, where
$k_F$ is the Fermi momentum of the non-interacting  fermions $c_a$. 
In a Luttinger liquid, the integers ($n,m$) are not arbitrary and depend on the boundary conditions. For instance, in a system with periodic boundary conditions, the total charge $Q$ and current $J$  carried by an excitation are such that $Q\pm J$ is even \footnote{This is due to the fact that only operators for which $m+n$ is even are expressible as a combination of electron and hole operators, and are hence local in space.}, which translates in the present case to 
\begin{equation}
n \pm mN \; {\rm even}, \; (n, m) \in \mathbb{Z}.\\
\label{eq: QJconstraint}
\end{equation}

We now observe that, due to (\ref{eq: conjugate}),  the vertex operator  $e^{i\pi \Phi(x)}$
introduces a $\pi$ kink in  the phase $\Theta$  and hence  from (\ref{eq: Thetatheta})
it creates $\pi$ kinks in the phases of all the flavors, $\theta_a$. We therefore conclude that switching 
the parity of all $N$ chains simultaneously results in inserting a current $N$.
We may now read off from \eqref{eq: QJconstraint} the nature of the massless excitations in the system.
When $N$ is even there are only bosonic excitations with even charges $ n=2p$. These excitations may or may not change the fermionic parity of all the flavors depending on the parity of $m$. When $N$ is odd, fermionic excitations with odd charges, e.g.  $n=2p+1$, necessarily change the fermionic parity of all the  flavors since  from (\ref{eq: QJconstraint}) $m$ is odd. In contrast,  bosonic excitations with even charges, e.g. $n=2p$, must be accompanied by an even number $m$  of $\pi$ kinks which do not change the fermionic parity of the flavors.
We thus recover the results discussed in \ref{subsec:N_flav}.

\subsubsection{$N$ odd and composite fermionic excitations}
When  $N$ is odd the excitations are either fermionic, for $n$ odd, or bosonic, for $n$ even.
The fundamental  excitation in this case is the charge $Q=1$ fermion created by (see Eq.~\eqref{eq:compositek})
\begin{equation}
\Psi_F^{\dagger}(x)  \propto c_{i, a_1}^{\dagger}... c_{i, a_{(N+1)/2}}^{\dagger} c_{i, b_1}^{}... c_{i, b_{(N-1)/2}}^{}.
\label{eq:compositefermion}
\end{equation}
This operator has a finite overlap with the vertex operators (\ref{eq:vertex}). 
For instance, to leading order we have
\begin{equation}
\Psi_F^{\dagger}(x)  \sim \Psi^{\dagger}_{L} \; e^{iNk_F x} + \Psi^{\dagger}_{R} \; e^{-iNk_F x} 
\label{eq:compositefermionboso}
\end{equation}
where  $\Psi^{\dagger}_{L/R}  \propto e^{i[ \Theta(x) \pm  \pi  \Phi(x)]}$ creates a charge one fermion with  left and right moving 
excitations at momenta $\pm Nk_F$. The charge-1 fermion (\ref{eq:compositefermion}) can hence be interpreted as a spinless fermion with an enlarged Fermi surface at $\pm Nk_F$. Furthermore, as discussed in \cite{AzariaLuttinger}, the system may be viewed as a fermionic Luttinger liquid made of interacting composite fermionic particles with Luttinger parameter $K$. In this respect notice that when $N=1$, the composite fermionic particle is identical to the bare fermion, and one recovers the usual Luttinger liquid description. 

At this point it is worth stressing that although, when $N$ is odd, the low-energy physics is described by interacting spinless charge-1 fermions, the generalized Luttinger liquid state for $N>1$ is different from that of a single flavor system. In particular, when expressed in terms of the lattice fermions, the quantum numbers of the low-energy excitations are different, as can be seen from (\ref{eq: QJconstraint}), as well as their momentum scale which is $Nk_F$. They do, however, share an essential feature in that there is no gap to Q=1 fermionic excitations. 
Hence, the superconducting character of the phase for $N$ odd is to be understood as a phase where the p-wave pair correlation function decays slower at large distances than 
any other correlation function, despite having no gap to adding single fermions. 

Consider for instance the pair creation operator
\begin{equation}
\Pi^{\dagger}_a(x) = c_{i, a}^{\dagger} c_{i+1, a}^{\dagger}  \propto e^{2i \Theta(x)}.
\label{eq:pwaveboso}
\end{equation}
As we have $\langle \Psi_F^{\dagger}(x) \Psi_F^{}(0) \rangle \sim x^{- (K+ K^{-1})/2}$ and $\langle \Pi^{\dagger}_a(x) \Pi^{}_a(0) \rangle \sim x^{- 2/K}$,  the pair correlations dominate for $K > 1/\sqrt{3}$.

Both the composite fermionic operator and the pair creation operators have sub-dominant components at the higher momenta $P=(2p+ 1) Nk_F$ and $P=2p Nk_F$. These are generated by the charge neutral  operator $e^{2i\pi \Phi(x)}$ which 
introduces  a $2\pi $-kink in the $\Theta$ field and hence does not change the boundary conditions of the $f$ fermions. The latter operator corresponds to the $N$'th moment of the density operator \cite{OshikawaAffleck}
\begin{eqnarray}
n_N(x)  &\sim& c_{i, 1}^{\dagger}... c_{i, N}^{\dagger} c_{i, 1}^{}... c_{i, N}^{} \nonumber \\
&\propto& e^{2i\pi \Phi(x) + 2iNk_F x} + {\rm h.c.}.
\label{eq:rhoN}
\end{eqnarray}
For instance,  the density operator $n(x)$, as any other bosonic operator,
has only momentum components at $2mNk_F$ and is given to  leading order by
\begin{equation}
n(x) \sim  \bar \rho +  \partial_x \Phi(x) + A_N e^{2i\pi \Phi(x) + 2iNk_F x} + {\rm h.c.},
\label{eq:densityodd}
\end{equation}
where $A_N$ is a non-universal constant. 

\subsubsection{$N$ even and composite bosonic excitations}

Given the constraint (\ref{eq: QJconstraint}), when $N$ is even,  $n$ has to be even as well, independently of $m$. 
Hence, there can only be bosonic excitations and there are no fermionic excitations in the low energy spectrum, contrary to the odd $N$ case.  To leading order, the elementary excitation in this case is given by the bosonic vertex operator 
\begin{equation}
\Psi_B^{\dagger}(x)   \propto e^{2i \Theta(x)}.
\label{eq:boso}
\end{equation}
 which has zero momentum. In terms of the lattice fermions, both 
 the charge $Q=2$ pair creation operator (\ref{eq:compositefermionboso}) and the charge $Q=2$ composite
 operator (\ref{eq:compositek}), with $M=N/2+1$, have a finite overlap with (\ref{eq:boso}). This stems from the fact
 that since  $N$ is even, the composite operators  (\ref{eq:compositek}) always introduce an even number of $\pm \pi$
 kinks in  the phases $\theta_a$ which may average to zero for the mean  $\Theta$ phase (\ref{eq: Thetatheta}).
Higher momentum corrections at $P=p Nk_F$ to (\ref{eq:boso}) are of course also generated. The charge neutral operator that generates them is $e^{i\pi \Phi(x)}$ which creates a $\pi$ kink in the phase $\Theta$. In terms of the 
fermions it is given by the composite density
\begin{eqnarray}
n_{\frac{N}{2}}(x)  &\sim& c_{i, a_1}^{\dagger}... c_{i, a_{N/2}}^{\dagger} c_{i, b_1}^{}... c_{i, b_{N/2}}^{} \nonumber \\
&\propto& e^{i\pi \Phi(x) + iNk_F x} + {\rm h.c.}.
\label{eq:rhoN/2}
\end{eqnarray}
We find, in particular, that the total density in this case has momenta components at $\pm Nk_F$
\begin{equation}
n(x) \sim \bar \rho + \partial_x \Phi(x) + B_N e^{i\pi \Phi(x) + iNk_F x} + {\rm h.c.}.
\label{eq:densityeven}
\end{equation}

\section{SO(N) symmetric coupled chains\label{sec:SON}}

We now turn to illustrate the principles we discussed above in a simple lattice model.
We shall consider the situation where an $SO(N)$ symmetry is present and set $V_{ij}^{ab}=V_{ij}=V\delta_{i,j}$ in Eq.~(\ref{HNflavors}). An on-site repulsive interaction will also be included. The Hamiltonian is given by
\bea
H&=&-t\sum_{i, a=1}^N (c_{i,a}^{\dagger}c_{i+1,a}^{} + {\rm h.c.})+ V\sum_{i}\Pi_{i}^{\dagger}\Pi_{i}^{} \nonumber \\
&+& \frac{U}{2}\sum_{i}\left(\sum_{a=1}^Nn_{i,a}\right)^{2},
\label{HSO(N)}
\eea
where the operator
\be
\Pi_{i}^{\dagger}=\sum_{a=1}^Nc_{i,a}^{\dagger}c_{i+1,a}^{\dagger}
\ee
creates a pair in a $SO(N)$ singlet state and  $n_{i,a} = c_{i,a}^{\dagger} c_{i,a}^{}$ is the density of the fermion of flavor $a$ at  site $i$. In the following we shall assume  $V<0$ and fix the filling to $\bar \rho=1/N$ for each flavor (one fermion per site). As we show below, the model (\ref{HSO(N)}) displays a rich
phase diagram, including both insulating and gapless phases, either with or without topologically protected edge modes.

In the following, we shall investigate in more details the physics associated with  (\ref{HSO(N)}) in the weak-coupling limit, i.e. $|U|/t \ll1$ and $|V|/t \ll 1$. In Sec. \ref{sec:GaplessHaldane} we will study numerically the case of $N=3$ in both the strong and the weak coupling limits.

\subsection{Field theory analysis\label{sec:Field-theory-analysis}}

In the weak-coupling limit, the low-energy physics associated with (\ref{HSO(N)}) 
is obtained in the  standard way by linearizing  the spectrum around the two
Fermi points $\pm k_F$ ($k_{F}=\pi \bar \rho$)  associated with the non-interacting fermions. The lattice fermions $c_{j, a}$ are expressed   
in terms of  left and right moving fermionic modes as

\begin{equation}
c_{j, a}\simeq \Psi_{L, a}e^{-ik_{F}x} + \Psi_{R, a}e^{ik_{F}x},
\end{equation}
where $x=j a_0$, $a_0$ being the lattice spacing, and $j$ is an integer.
The next step is to bosonize the fermions, writing $\Psi_{L(R),a}$ as 
\begin{equation}
\Psi_{L\left(R\right), a}\left(x\right)=\frac{\kappa_{a}}{\sqrt{2\pi}}e^{ -i \left[\theta_a(x) \pm \pi
\phi_a\left(x\right)\right]}, 
\label{bosonization}
\end{equation}
where $\kappa_{a}$ are  Klein factors satisfying  $\left\{ \kappa_{a},\kappa_{b}\right\} =2\delta_{a, b}$ to  ensure fermionic
anti-commutation relations. The bosonic fields $\phi_{a}\left(x\right)$ and $\theta_{b}\left(x\right)$ satisfy the equal-time
commutation relations $[ \phi_{a}\left(x\right), \theta_{b}\left(y\right)]= i \delta_{a, b}  Y(y-x)$
and are related to the current density, $j_a\left(x\right) = \partial_x \theta_a\left(x\right)/\pi $, and to the density of each flavor
\begin{equation}
n_{a}\left(x\right) \simeq \bar{\rho}_{}+\partial_{x}\phi_{a}\left(x\right)+\frac{i}{2\pi}e^{2i \pi \phi_{a}\left(x\right)+ 2ik_{F}x}+h.c. .
\label{densityboso}
\end{equation}
At this point it is convenient to perform a change of basis to the bosonic fields
$\left(\Phi_{},\vec{\Phi}\right)$ and $\left(\Theta_{},\vec{\Theta}\right)$
corresponding to a collective (or charge) mode and to $N-1$ spin modes as follows
\bea
\Phi_{}&=& \sum_{a=1}^{N}\phi_{a},\ \ \ \  \vec{\Phi}=\sum_{a=1}^{N}\vec{\omega}_{a}\phi_{a},
\nonumber \\
\Theta_{}&=& \frac{1}{N} \sum_{a=1}^{N}\theta_{a},\ \vec{\Theta}=\sum_{a=1}^{N}\vec{\omega}_{a}\theta_{a},
\label{eq:phi_trans} 
\eea
where $\vec{\omega}_{a=1,...,N}$ are $N-1$ components vectors
satisfying $\vec{\omega}_{a}\cdot\vec{\omega}_{b}=\delta_{a, b}-1/N$ 
and  $\sum_{a=1}^N \vec{\omega}_{a}=0$. The latter conditions ensure that
the transformation (\ref{eq:phi_trans}) is canonical. For completeness
we write also the  inverse transformation
\bea
\phi_{a}&=&\Phi_{}/N+\vec{\omega}_{a}\cdot\vec{\Phi}, \nonumber \\
\theta_{a}&=&\Theta_{}+\vec{\omega}_{a}\cdot\vec{\Theta}.
\label{eq:phi_transinv} 
\eea

Using these definitions, we find that the  low-energy physics
of our model  is described by  the following bosonized Hamiltonian 
with decoupled spin and charge sectors, $H=H_{c}+H_{s}$, where

\begin{equation}
H_c=\frac{v_{c}}{2\pi }\int dx  [K_{}\left(\partial_{x}\Theta_{}\right)^{2}+\frac{\pi^2}{K_{}}\left(\partial_{x}\Phi_{}\right)^{2} ],
\label{LuttWC}
\end{equation}
and
\bea
&H_s& =\int dx \{  \frac{v_{F}}{2\pi} [ (\partial_{x}\vec{\Theta} )^{2}+\pi^2 (\partial_{x}\vec{\Phi})^{2} ]+\frac{g_{\perp}}{2} (\partial_{x}\vec{\Phi} )^{2} \nonumber \\ 
 &-&\frac{1}{2\pi^{2} }\sum_{a< b}[ \lambda  \cos (2 \pi \vec \alpha_{ab}\cdot \vec \Phi)+ \tilde \lambda \cos (2 \vec \alpha_{ab} \cdot\vec \Theta ) ] \}.  \nonumber \\ 
 \label{eq:Hbos}
\eea
In Eq.~\eqref{LuttWC}, $v_{F}= \pi \bar \rho$ is the Fermi velocity of the non-interacting fermions, $K_{}=N/\sqrt{1+\frac{\left(N-1\right)U+\bar{V}}{\pi v_{F}}}$, where $\bar{V}=2V\left(1-\cos\left(2k_{F}\right)\right)$, is the charge Luttinger parameter and   $v_{c}=v_{F}/K$ is the charge velocity. In the spin sector (\ref{eq:Hbos}), $\vec \alpha_{ab}= \vec \omega_a - \vec \omega_b$ and   the couplings, $\lambda =(g_{\parallel} + g_{\perp})/2$ and its dual $\tilde \lambda =(g_{\parallel} - g_{\perp})/2$,  are related to  $U$ and $\bar V$ by  $g_{\parallel,\perp}=-U\mp\bar{V}$.

\subsubsection{Spin Sector}

Let us first discuss the spin sector. The Hamiltonian (\ref{eq:Hbos}) 
describes the competition between two mutually incompatible ordering tendencies
favored by the two cosine terms in  (\ref{eq:Hbos}).  We thus expect that either the $\vec \Phi$ field  or its dual  $\vec \Theta$  gets locked when these terms are relevant. The phase diagram in the spin sector results from a delicate balance between the three interaction terms entering (\ref{eq:Hbos}). 
The RG equation associated with the couplings in (\ref{eq:Hbos}) are given to the one-loop order by \cite{Itoi1997}
\bea
\frac{d\tilde{g}_{\parallel}}{dt} & = & \left(N-2\right)\tilde{g}_{\parallel}^{2}+\left(N+2\right)\tilde{g}_{\perp}^{2},\\
\frac{d\tilde{g}_{\perp}}{dt} & = & 2N\tilde{g}_{\parallel}\tilde{g}_{\perp},
\eea
where we have rescaled the couplings as $\tilde{g}_{\parallel,\perp} = g_{\parallel,\perp}/4\pi v_{F}$.

For an attractive pairing interaction, i.e. when $\bar V  < 0$,  the RG flow drives the system towards strong coupling
and a spin gap opens. The nature of the resulting phase depends on the relative strength of
the on-site interaction $U$ and the pairing term. We distinguish between two phases.

\paragraph{$SU(N)$ Phase.\label{sunphase}}

When $U < \bar V  < 0$, i.e. when the  on-site attraction dominate the  pairing term, 
the RG flow drives the system toward the attractive ray: $g_{\parallel}=g_{\perp} = \lambda > 0$. On this line the interacting part of (\ref{eq:Hbos}) takes the form
\be
H_{\rm int}= \frac{\lambda}{2} \int dx \;  [ (\partial_{x}\vec{\Phi} )^{2} 
 - \frac{1}{\pi^{2} }\sum_{a< b}  \cos (2 \pi \vec \alpha_{ab}\cdot \vec \Phi)].
 \label{HSUN}
\ee
When $\lambda = g_{\perp} > 0$, the interaction is relevant and  
a spin gap,  $\Delta_{\sigma} \propto e^{-1/2N\lambda}$, opens.
The spin field $\vec \Phi$ gets locked in such a way that the cosine terms in (\ref{HSUN})
are maximal. Using (\ref{eq:phi_transinv}) we find that $\vec \alpha_{ab}\cdot \vec \Phi = \phi_a - \phi_b$,
which  implies for the  flavor fields: $\phi_a - \phi_b = p$, $p \in \mathbb{Z}$. As a consequence the fluctuations of the different flavor densities, $n_a(x)$ in  (\ref{densityboso}), are in phase with each other.
In this phase $N$ fermions of different flavors bind together into charge $Q=N$ spin-singlets, forming a compressible fluid. In particular, for $N=3$, this is the trionic phase discussed in \cite{LecheminantTrions} in which triplets of fermions bind together. This  phase is ``topologically trivial" (i.e. it has no protected edge states)  since it is adiabatically connected to a phase, obtained when $U=-\infty$, where the $N$-fermion bound states can be described essentially as free spinless fermions. 

Although it is not obvious when written in terms of the spin field  $\vec \Phi$, the Hamiltonian (\ref{HSUN}) possesses an emergent  $SU(N)$ symmetry. To see this, observe that  this fixed point corresponds to the pairing $V$ flowing to zero at which the lattice Hamiltonian (\ref{HSO(N)}) is clearly $SU(N)$ invariant. Therefore in the whole domain $U < \bar V  < 0$ an $SU(N)$ symmetry is dynamically enlarged in the low-energy limit.

\paragraph{Dual $\widehat{SU(N)}$ Phase.\label{dualsunphase}}

When the pairing term dominates the physics, e.g. when $|\bar V|  > |U|$, 
 the RG drives the system toward the attractive ray: $g_{\parallel}=- g_{\perp} = \lambda > 0$, 
 at which the interacting part of (\ref{eq:Hbos}) takes the form
\be
H_{\rm int}= \frac{\tilde\lambda}{2 \pi^2}  \int dx \;  [ (\partial_{x}\vec{\Theta} )^{2} 
 - \sum_{a< b}  \cos (2  \vec \alpha_{ab}\cdot \vec \Theta)].
 \label{HSUNdual}
\ee
As in the previous phase a spin gap, $\Delta_{\sigma} \propto e^{-1/(2N\tilde \lambda)}$, opens but this time it is the
 $\vec \Theta$ field that gets locked instead of $\vec \Phi$. From (\ref{eq:phi_transinv}), we find that  this implies:  $\theta_a- \theta_b = p \pi$, $p \in \mathbb{Z}$, exactly as in the gapless topological phase discussed in Sec. \ref{subsec:N_flav}. We therefore expect that in the dual $\widehat{SU(N)}$ phase the system host 
 protected zero-energy edge states. These edge states transform as a spinor representation of $SO(N)$, as discussed in Sec.~\ref{sec:endmodes}.

The fixed points Hamiltonians (\ref{HSUN}) and (\ref{HSUNdual}) are related by the duality transformation
\be
\Omega: \vec{\Theta} \rightleftarrows \pi \vec{\Phi}, \; \Omega^2=1.
\label{duality}
\ee
We therefore find that the massless topological phase is \emph{dual} in the sense of 
(\ref{duality}) to the topologically trivial $SU(N)$ phase. As discussed in \cite{BoulatDuality}
 the $SU(N)$ symmetry of (\ref{HSUN}) translates by the duality (\ref{duality})
 into a dual  $\widehat{SU(N)}$ symmetry for (\ref{HSUNdual}) \footnote{In the terminology used in 
\cite{BoulatDuality}  the  dual  fixed point Hamiltonian (\ref{HSUN}) belongs to the class $\cal{A}_{I}$
 and is  is associated with the $SO(N)$ symmetry of the problem.}.

\paragraph{Self-Dual Phase Transition Line.\label{selfdualline}}
When $\bar V  =U$ there is a quantum phase transition between the two phases. At this point,
 $g_{\perp}=0$, $g_{\parallel} = \lambda  >0$, and the interacting part 
of Hamiltonian (\ref{eq:Hbos}) takes the form
\be
H_{\rm int}= \frac{- \lambda}{2 \pi^2}  \int dx \;  [ \sum_{a< b}  \cos (2 \pi \vec \alpha_{ab}\cdot \vec \Phi) +\cos (2  \vec \alpha_{ab}\cdot \vec \Theta)],
 \label{HSUNselfdual}
\ee
which is invariant under the duality transformation (\ref{duality}). 
This is the self-dual line, separating the two dual $SU(N)$ and 
$\widehat{SU(N)}$ symmetric phases discussed above. 
As shown by Andrei and Destri \cite{Andrei1984},  the model (\ref{HSUNselfdual})
is integrable and a spin gap $\Delta_{\sigma} \propto e^{-1/(N-2)\tilde \lambda}$
is still present (notice  that the spin gap on the self-dual line is  parametrically 
smaller than in both  $SU(N)$ and   $\widehat{SU(N)}$  phases).
This suggests that the phase transition between the trivial $SU(N)$ phase and the dual 
$\widehat{SU(N)}$ topological phase  is of first order. The model (\ref{HSUNselfdual}) undergoes
 dynamical symmetry breaking of the dual symmetry (\ref{duality}) with kinks that carry zero modes transforming according 
 to the spinor representation of $SO(N)$. In the particular case of $N=3$
the model hosts propagating spin-$1/2$ kinks. This allows us to interpret these spinor kinks   as the end-modes hosted by the dynamical domain  walls between the trivial and topological phases. We will address this result further in Sec. \ref{sec:GaplessHaldane}.

\subsubsection{Charge Sector}

In all phases discussed above the low-energy sector is described by the generalized Luttinger liquid 
Hamiltonian (\ref{LuttWC}). As far as charge excitations are concerned,
the low-energy physics depends only on the non-universal Luttinger parameters
$K(U,V)$ and $v_c(U,V)$. However, the low-energy excitations in the $SU(N)$ and $\widehat{SU(N)}$ symmetric 
phases have a different character. Their nature is encoded 
in the total charge $Q$ and total current $J$ (or zero-mode) spectrum of (\ref{LuttWC}).

In the gapless topological phase, as discussed in section \ref{sec:lowenergy},
we found that, due to the topological nature of a single Kitaev chain and the
locking of the spin field $\vec \Theta$, the low-energy excitations either
change the parity of all the flavors simultaneously or do not change 
them at all. We then deduced that the total charge and current  in the system 
were given by (\ref{QJ})
\be
 Q=n, J=mN, \; Q\pm J \; {\rm even},
 \ee
where $(n,m)\in \mathbb{Z}$. The fundamental excitation is  either
a charge $Q=2$ boson for $N$ even or a composite fermion of charge $Q=1$
for odd $N$ \cite{AzariaLuttinger}. In both cases the Fermi momentum $P_F$ is enlarged to $P_F=Nk_F/2$
and $P_F=Nk_F$ respectively.

As seen above, the $SU(N)$ symmetric topologically trivial phase was obtained from 
the topological $\widehat{SU(N)}$ symmetric one by the duality transformation  (\ref{duality}) 
on the spin bosonic fields $\vec \Phi$ and $\vec \Theta$. In the charge sector the latter
duality translates onto the charge fields
\bea
  N \Theta \rightleftarrows \pi \Phi.
 \eea
We may then deduce the zero mode  spectrum of the Luttinger liquid Hamiltonian (\ref{LuttWC}) 
in this phase
\be
 Q=nN , J=m, \; Q\pm J \; {\rm even}.
\ee
We immediately see that   it is the total charge of the system, instead of the current, that is quantized in units of $N$.
The fundamental excitations are $SU(N)$ singlet bound-states made
of $N$ fermions. They are bosons for $N$ even and fermions for $N$
odd. In both cases the Fermi momentum is that of the lattice fermions
e.g. $P_F=k_F$ \footnote{In the classification of Luttinger liquid proposed
in \cite{AzariaLuttinger} these two phases belong to the classes ${\cal A}_I$
and ${\cal A}_0$.}.

\section{The gapless Haldane phase in SO(3) symmetric fermion chains\label{sec:GaplessHaldane}}

In order to illustrate the ideas we laid out above, we now study the model \eqref{HSO(N)} in more detail, focusing on the
case of $N=3$. We consider the system to be at $1/3$ filling (one
fermion per site), and start by discussing the different limits of
the model, and the phases that are expected to arise in those limits. 
A quantitative phase diagram,
obtained using DMRG, is presented in Fig. \ref{fig:PD} and will be
described in greater detail in Sec. \ref{subsec:Phase-diagram}. 

We first consider the case of repulsive on-site interactions, $U>0$,
and vanishing pairing interactions, $V=0$. In this case, the Hamiltonian
has an $SU\left(3\right)$ symmetry. In the strong coupling limit, $U\gg t$, 
we expect the system to be in a Mott phase with a gapped charge sector \cite{RolandAzaria1999}.
The low-energy effective Hamiltonian, obtained by second-order perturbation
theory in $t/U$, is then given by the bilinear-biquadratic spin-$1$ model
\begin{equation}
H=\sum_{i}\left(J_{1}\vec{S}_{i}\cdot\vec{S}_{i+1}+J_{2}\left(\vec{S}_{i}\cdot\vec{S}_{i+1}\right)^{2}\right),\label{eq:HS}
\end{equation}
where $S_i^a=i\epsilon^{a b c}c_{i,b}^{\dagger}c_{i,c}$
are $S=1$ operators, and $J_{1,2}=J=2t^{2}/U$. 
The Hamiltonian \eqref{eq:HS} was studied extensively (see, e.g.,~\cite{Schollwock1996,HongHao2008} and the references therein), and exhibits a rich phase diagram as a function of $J_2/J_1$. In particular, for $J_1=J_2$, the model is critical and described by a level-one $SU\left(3\right)$ Wess-Zumino-Witten model \cite{Itoi1997}.

When a nonzero $V$ is introduced, the symmetry of the model is reduced
to $SO\left(3\right)$. In the limit $U\gg t$ and for $V\sim t^{2}/U$,
the low-energy effective Hamiltonian is given by \eqref{eq:HS} with $J_{1}=J$
and $J_{2}=J+V$. For $-2J<V<0$ the system is fully gapped, and belongs
to the Haldane phase with decoupled spin-$1/2$ degrees of freedom
at its ends, similarly to the AKLT model \cite{AKLT1987}. For $V<-2J$
a dimerized phase, which breaks translational invariance, is expected. 

Going back to the weak coupling limit, i.e. $|U|,|V|\ll t$, 
we recall the analysis presented in Sec. \ref{sec:Field-theory-analysis}, 
which suggested that for $U>0$ and $V<0$ the system is in the gapless topological phase.
To understand the nature of the topological phase and the end-modes in this case, 
recall that the mean-field description of this phase hosts three Majorana zero modes at each end of the chain, $\gamma_{L(R), a}$.
These modes can be combined to form a spin-$1/2$ degree of freedom 
$S_{L(R)}^a= -\frac{i}{4}  \epsilon^{a b c} \gamma_{L(R), b} \gamma_{L(R), c}$, 
similarly to the spin-$1/2$ degrees of freedom in the Haldane phase.
In the gapless topological phase, these end-modes remain localized at the two ends of the system, and cannot couple to one another due to the spin gap in the bulk.
Due to the close relation with the Haldane phase, we refer to this phase as the gapless Haldane phase.

Finally, for attractive on-site interactions, $U<0$, we expect to find a phase transition
into the trivial, trionic phase, in which triplets of fermions of
different flavors bind together forming an $SU\left(3\right)$ singlet. 

\subsection{Phase diagram\label{subsec:Phase-diagram}}

\begin{figure}[!h]
\includegraphics[bb=0bp 0bp 720bp 440bp,clip,width=1\columnwidth]{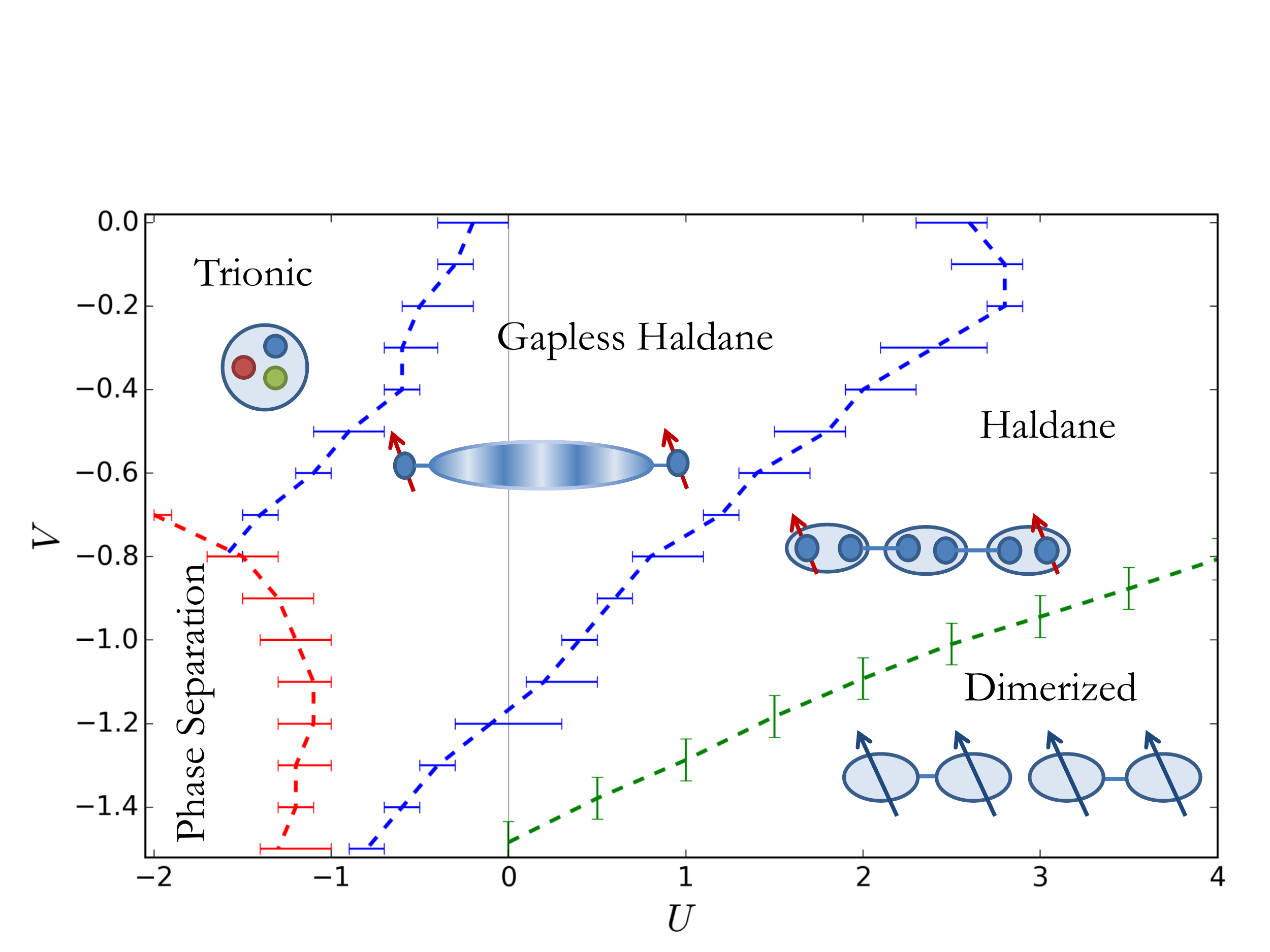}

\caption{Phase diagram of the model \eqref{HSO(N)} with $N=3$ at $1/3$ filling,
as function of the on-site interaction $U$ and nearest neighbor pairing interaction
$V<0$. For repulsive on-site interactions, $U>0$, a charge gap opens
as $U$ is increased and the system goes from the gapless to the
gapped Haldane phase. For large enough $\left|V\right|$ a transition to the fully
gapped dimerized phase is observed. For attractive on-site interactions,
$U<0$, the system undergoes a transition into the trionic phase,
in which triplets of fermions bind together into spin-singlets. For
large attractive $U$ and $V$ the system tends to phase separate.
For details on how the phase boundaries are determined, see main text
and Appendices \ref{appx:ChargePB}, \ref{appx:DimerPB}.\label{fig:PD}}
\end{figure}

We describe below the phase diagram of the model, as function
of the on-site interaction strength $U$, and the nearest neighbor
pairing $V<0$, obtained using DMRG \cite{ITensor} 
and presented in Fig. \ref{fig:PD}. 
In the DMRG calculation, we represent each fermionic flavor as a single chain, 
and work in the basis $c_{\pm1}=(c_1\pm i c_2)/\sqrt{2}$, $c_0=c_3$, such that $S^3=n_{+1}-n_{-1}$ is conserved (hereafter, we will denote $S^3$ by $S^z$).

In agreement with the weak coupling analysis presented
in Sec. \ref{sec:Field-theory-analysis}, we find that as $\left|V\right|$
is increased, a finite spin gap opens in the system (see Appendix
\ref{appx:SpinGap} for more details). Hence, all the phases discussed
below have a fully gapped spin sector.

\subsubsection{Mott transition}

As discussed above, for large enough $U>0$ we expect the system to
undergo a Mott transition as a charge gap opens. For small $V<0$
the charge gap opening line separates the gapped and gapless Haldane
phases. To see where this phase transition occurs, we calculate the
gap to adding a single particle. 
{
For a fixed system size $N_x$ we calculate
\begin{equation}
E_{3n=N_x}^{{\rm SP}}=E_{3n+1}+E_{3n-1}-2E_{3n},\label{eq:E_SP}
\end{equation}
where by $E_n$ we denote the ground state energy of a system
with $n$ particles.} 
We then extrapolate $E_{3n=N_x}^{{\rm SP}}$ to the infinite system
size limit. Indeed, we find a finite region in the parameters space,
in which for $U<U_{c}\left(V\right)$ the charge sector is gapless.
For more details on how $U_{c}\left(V\right)$ is determined see Appendix
\ref{appx:ChargePB}.

\subsubsection{Dimerization transition}

The strong coupling arguments presented above also suggest that large
attractive pairing interactions drive a transition into a dimerized phase. 
The open boundary conditions used in the simulations always induce some dimerization in the middle region of the system.
To identify the transition into the dimerized phase, we calculate
the local dimerization $D_{i}=\left|\vec{S}_{i}\cdot\vec{S}_{i+1}-\vec{S}_{i-1}\cdot\vec{S}_{i}\right|$. 
At the dimerization transition, the dimerization in the middle
of the chain is expected to decay as a power law $D_{N_{x}/2}\sim N_{x}^{-d}$
with an exponent $d=3/8$ \cite{Chepiga2016}.

Performing finite size scaling, we fit $D_{N_{x}/2}$ to the form
given above, extracting the exponent $d$ as function of $V$, and
identify the phase transition point as the value of $V$ for which
the exponent equals $3/8$. For further details and numerical results
see Appendix \ref{appx:DimerPB}. 

\subsubsection{Trionic phase transition}

For large $U<0$, we expect to find a trionic phase, in which triplets
of fermions of different flavors bind together. Although the charge
sector in this phase is gapless, similarly to the gapless Haldane
phase, the two phases have a qualitatively different spectrum: 
the trionic phase has a finite gap to one and two fermion excitations, 
whereas in the Haldane phase both excitations are gapless. 
In addition, as we explain below, the two phases are topologically distinct. 
To obtain the phase transition line between
these two phases we once again calculate the single particle gap defined
in \eqref{eq:E_SP}, extrapolating it to the infinite system size
limit. 

From the field theoretical arguments (see Sec. \ref{sec:SON} and Ref. \cite{Andrei1984}), the phase transition between
these two phases is expected to be first order. 
However, while the distinct behaviors in the two phases are readily verified, we find it
difficult to verify the nature of the phase transition numerically,
due to the small size of the spin gap at the transition point. (Note
that the smallness of the gap at the transition point is consistent
with the weak coupling analysis above. See Appendix \ref{appx:TrionicPT}
for results and further discussion).

\subsubsection{Phase separation}

For large $U,V<0$ the system tends to phase separate, forming clusters
of trions. Interestingly, close to the line of the phase transition
between the gapless Haldane and the trionic phase the system phase
separates into regions in the gapless Haldane phase and a clustered
trions region, with localized spin-$1/2$ modes at the boundaries,
in support of the phase transition between the gapless Haldane and
the trionic phases being first order (see Appendix \ref{appx:TrionicPT} and Fig. \ref{fig:phase_sep} therein).

\subsection{Gapless Haldane phase\label{subsec:Gapless-Haldane-phase}}

\begin{figure}
\subfloat[\label{fig:dE}]{\includegraphics[width=0.5\columnwidth]{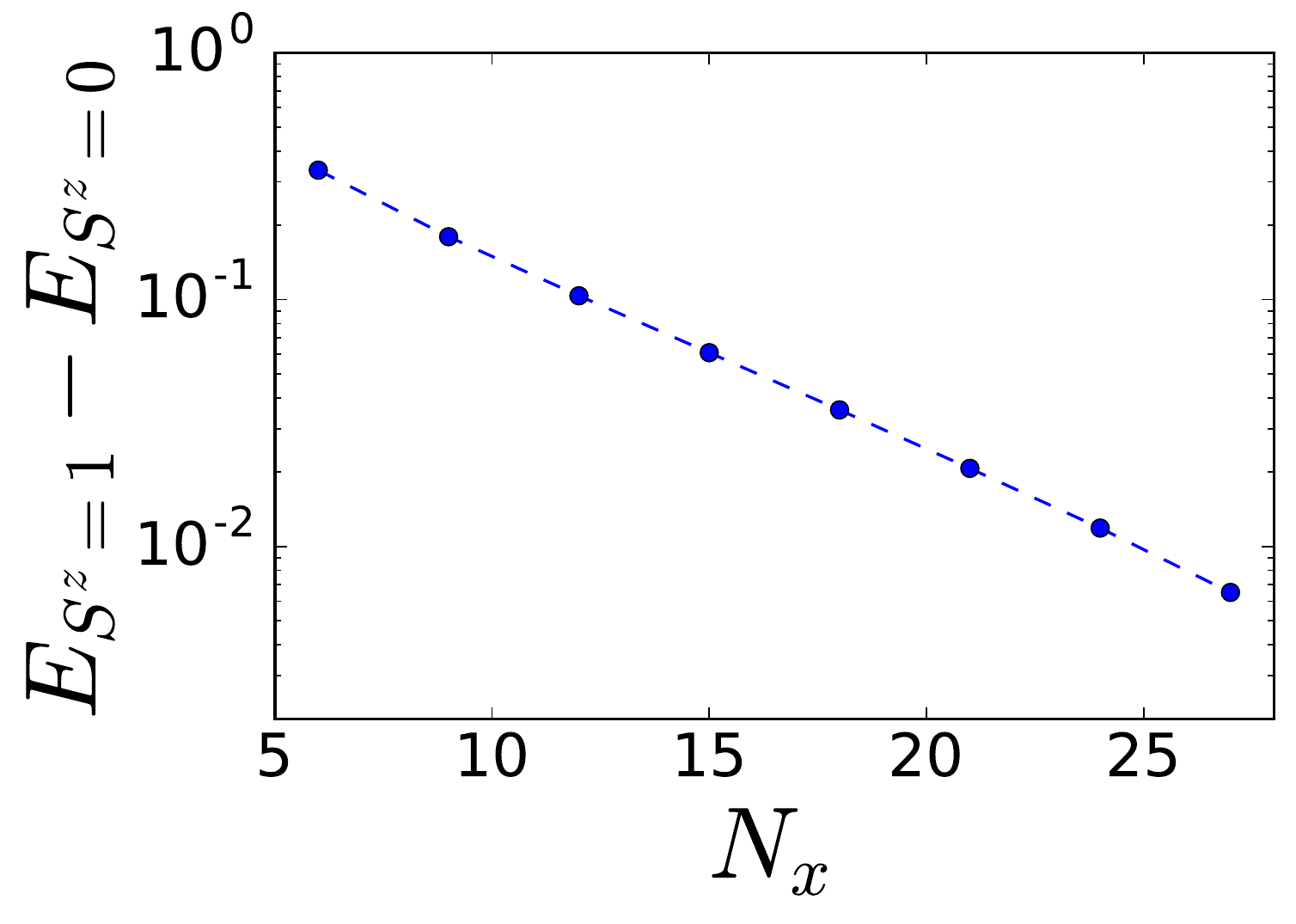}

}\subfloat[\label{fig:Sz}]{\includegraphics[width=0.5\columnwidth]{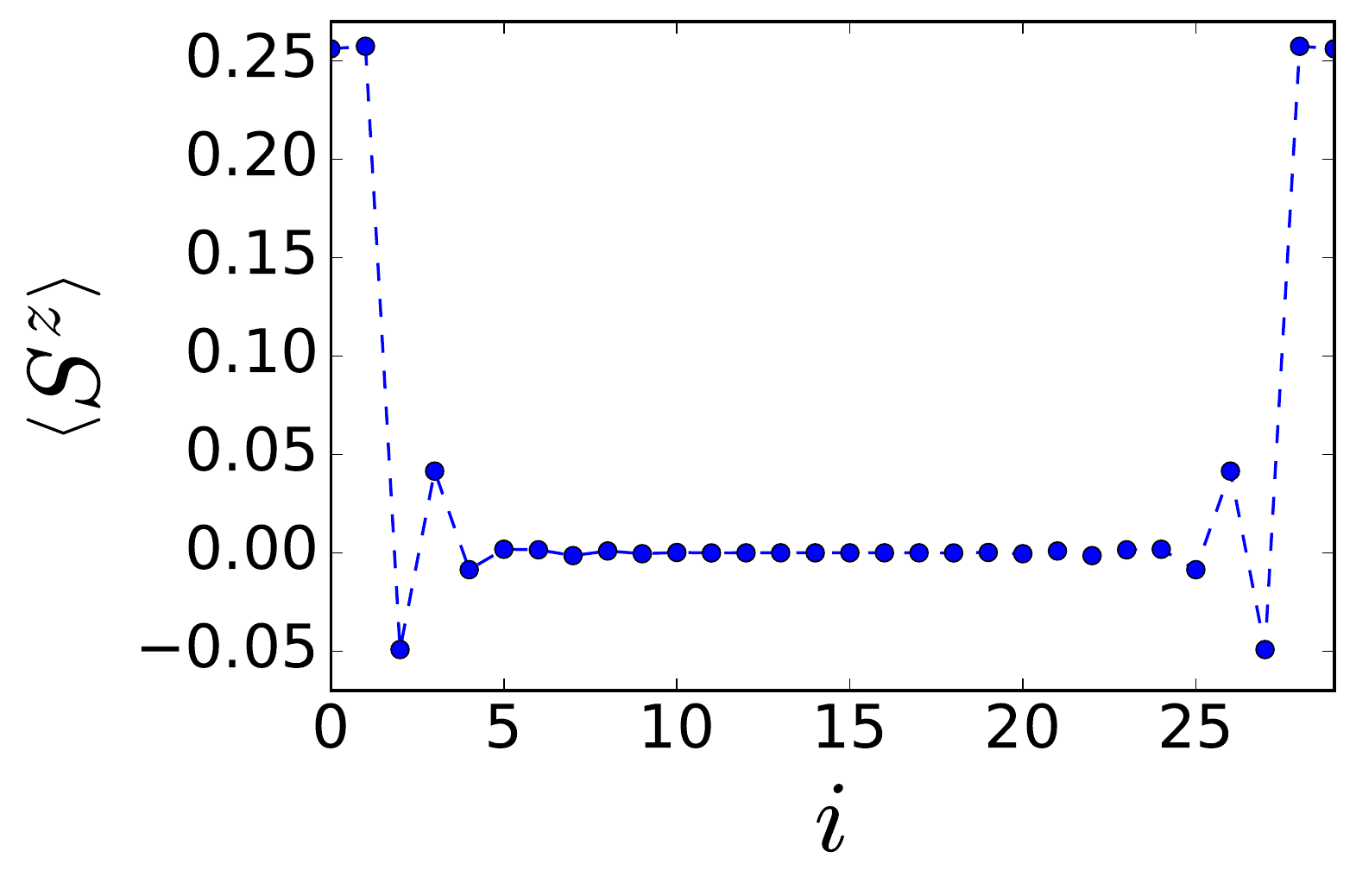}}

\subfloat[\label{fig:Two-point-correlations}]{\includegraphics[width=1\columnwidth]{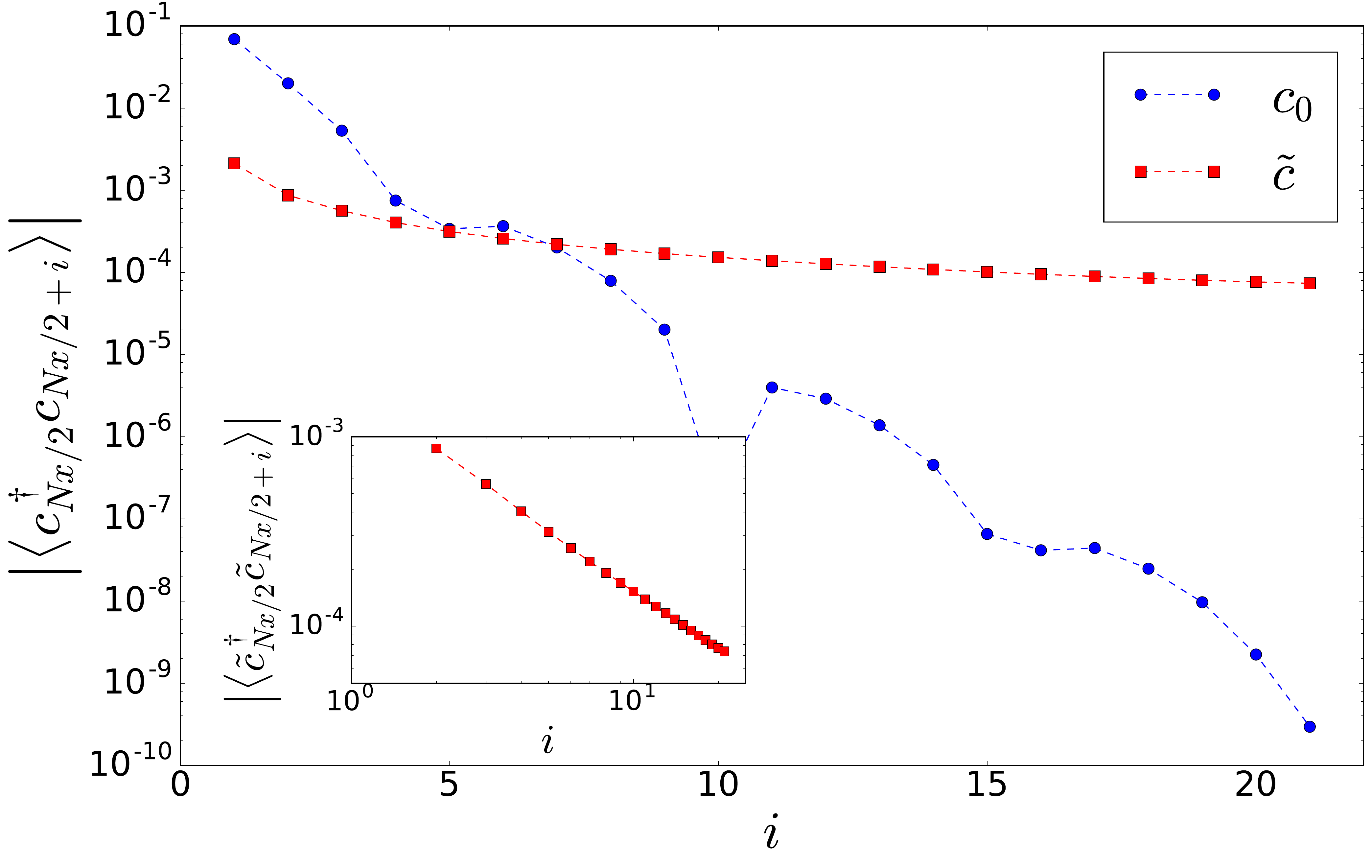}

}

\caption{(a) Energy splitting between the ground states with total $S^{z}=1$
and total $S^{z}=0$ as function of the length of the system for $U=0,V=-0.5$
shown on a semi-log scale. (b) The expectation value of $S^{z}$ as
function of position along the chain in the ground states of the system
of length $N_{x}=30$ sites, model parameters $U=0$, $V=-0.8$, and
total $S^{z}=1$. Exponentially localized spin-$\frac{1}{2}$ degrees
of freedom can be observed at the ends of the system. (c) Two-point
correlations of the bare fermions $\left\langle c_{0}^{\dagger}\left(i_{0}\right)c_{0}\left(i_{0}+i\right)\right\rangle $,
and of the charge-1 spinless fermions $\left\langle \tilde{c}^{\dagger}\left(i_{0}\right)\tilde{c}\left(i_{0}+i\right)\right\rangle $,
where $\tilde{c}^{\dagger}=c_{0}c_{1}^{\dagger}c_{-1}^{\dagger}$,
plotted in blue (circles) and red (squares) respectively, calculated
in the bulk of the system $i_{0}=N_{x}/2$ for a system of size $N_{x}=66$
sites and model parameters $U=0,V=-0.8$. The main plot shows the
correlations on a semi-log scale, while the inset shows the correlations
for the charge-1 spinless fermions on a log-log scale. It can be clearly seen
that while the former are exponential, the latter are power law as
expected.}

\end{figure}

We now further analyze the region in the phase diagram that we identified
as the gapless Haldane phase and discuss its properties. 
{
The edges of this phase are expected to support spin-$1/2$ end modes that can pair into a spin-singlet with $S^z=0$ or a spin-triplet with $S^z=0,\pm1$.
To show the exponential protection of the end-modes, we calculate the ground state energy in the total $S^{z}=0$
and $S^{z}=1$ sectors. As can be seen in Fig.~\ref{fig:dE}, this energy splitting decays exponentially in system size.
Furthermore, we calculate the local expectation value of $S^{z}$ in the ground state of the system in the total $S^z=1$ sector (see Fig. \ref{fig:Sz}). 
One can clearly see the localized spins at the two ends of the system (we have checked that $S^z$ near each edge indeed sums to $1/2$).
}

From the discussion in Sec. \ref{sec:StabilityTSC}, we expect the
bare fermions of each flavor to be gapped in the bulk, even though
the charge sector is gapless. To see this explicitly for our model,
we calculate the two-point correlation function of the fermions
(e.g. the $c_{0}$ fermions) in the bulk, $\left\langle c_{0}^{\dagger}\left(0\right)c_{0}\left(x\right)\right\rangle $,
and find that it decays exponentially with $x$ (see Fig. \ref{fig:Two-point-correlations}).
Similarly, we consider the correlations of the
the charge-1, spinless fermionic operator $\tilde{c} = c_{0}^{\dagger}c_{1}c_{-1}$,
$\left\langle \tilde{c}^{\dagger}\left(0\right)\tilde{c}\left(x\right)\right\rangle $. This operator changes the fermion parity of all three fermion flavors; therefore, by the arguments of Sec.~\ref{sec:StabilityTSC}, we expect it to have power law correlations in the gapless Haldane phase. 
These are also plotted in Fig. \ref{fig:Two-point-correlations}.
It can be seen that these correlations decay as a power law, with
an exponent close to unity. 
For the ease of presentation, we plot the absolute value of the correlations in Fig. \ref{fig:Two-point-correlations}. We note, however, that the correlations of the $\tilde{c}$ fermions exhibit $3k_F=\pi$ oscillations as expected from the low energy analysis presented in Sec. \ref{sec:lowenergy}.

At this point a natural question to ask is whether the gapless Haldane phase is stable to breaking of the $SO\left(3\right)$ symmetry. 
Given our general heuristic argument we expect the phase to be stable to any perturbation which preserves the $(\mathbb{Z}_2)^3$ symmetry, associated with the 
conservation of the fermionic parities of the three species. For instance, we expect that if a small single-ion anisotropy term, e.g.  $D_{z}\sum_i \left(S^{z}_i\right)^{2} = D_{z} \sum_i \left(n_{ i, +1} - n_{i, -1}\right)^{2}$, is added to the Hamiltonian~\eqref{HSO(N)}, the induced coupling between the end-modes will be exponential in system size. In Appendix \ref{appx:SIA} we present DMRG results, which are consistent with the latter expectation.

\section{Conclusions}
\label{sec:conclusions}

In this work, we have studied charge-conserving one-dimensional superconductors
with generic intrinsic attractive pairing interactions.
We  presented a general heuristic argument that implies that 
a system of spinless electrons with no additional symmetries does not support a topological phase with exponentially localized edge states.
In contrast, in the presence of additional symmetries, there are distinct topological superconducting phases with exponentially localized end modes. 
The properties of the end modes are directly related to those of a corresponding model with explicit mean-field superconducting terms, where the total charge is not conserved. 

We examined in detail a situation with
$N>1$ flavors of fermions, in which the fermionic parity of each flavor is separately conserved, resulting in a $(\mathbb{Z}_2)^N$ symmetry.  
In this case, the system hosts exponentially localized low-energy end-modes, which
are reminiscent of the Majorana zero-energy bound-states found in proximity coupled systems.
The stability of these edge states is  ensured  by the presence of a ``flavor gap''
due to the attractive pairing interactions 
and by the very special nature of the 
gapless modes of the system. Indeed, due to the $(\mathbb{Z}_2)^N$ symmetry we showed that the 
gapless modes either do not change the fermionic parity of any of the flavors or change them all simultaneously. 
As a consequence, the coupling between the edges through the gapless bulk modes only lifts the degeneracy between even and odd {\it total} fermionic 
parity sectors. The resulting topological degeneracy of the system is $2^N-1$.
We thus found a connection between the nature of the {\it bulk} low-energy excitations and the
existence of a topological phase in a charge conserving system. In particular, these excitations are created by composite
bosonic operators with an even charge for $N$ even, and composite fermionic operators with an odd charge and an enlarged Fermi surface at $\pm Nk_F$,
for $N$ odd. The analysis is readily generalized to other symmetry classes; we discussed the case of a single flavor with time reversal symmetry as an example. 

In order to be more explicit, we introduced and studied a simple paradigmatic 
model which, we believe, captures the essential features of charge conserving
topological superconductors. It includes both on-site Hubbard interactions and nearest-neighbor attraction,
and has an enlarged $SO(N)$ symmetry. Using field theoretical techniques, we showed that a phase which hosts 
massless composite bosons and fermions with a large Fermi surface is stabilized. 
Using the general arguments, we showed that this phase supports zero-energy bound-states
at its edges, which transform as $SO(N)$ spinors.
This gapless topological phase is separated from the trivial phase obtained for a large attractive Hubbard coupling by a quantum phase transition.  

Interestingly, the topological phase in the $N=3$ case 
is nothing but a gapless analogue of the Haldane phase of $S=1$ spin chains, where the three Majorana zero modes of the topological superconductor can be identified with the localized spin-$1/2$ edge states of the Haldane phase. This phase has gapless charge excitations in the bulk (including fermionic excitations with a unit charge). The edge states are protected either by the $SO(3)$ symmetry, or more generally by the separate conservation of the parity of each of the three fermion flavors. Upon opening a charge gap, this phase becomes the conventional (insulating) $S=1$ spin chain. 

Finally, we identify the transition between the gapless Haldane phase and the trivial trionic phase with the self-dual transition found in an integrable model by Andrei and Destri~\cite{Andrei1984}. This naturally explains the degeneracy associated with the kinks between the two phases, that was noted in~\cite{Andrei1984}, as domain walls between two topologically distinct phases that carry spin-$1/2$ zero modes. 
\vspace{5mm}

\acknowledgements
E.B. acknowledges support from the European Research
Council (ERC) under the European Union Horizon
2020 Research and Innovation Programme (Grant
Agreement No. 639172), from a Minerva ARCHES prize, and from CRC 183 of the Deutsche Forschungsgemeinschaft. We acknowledge the hospitality of KITP at UCSB,
which is supported by NSF grant PHY-1125915.
P.A. wants to thank the Weizmann Institute for its kind hospitality while this work started.

\bibliography{ref}

\begin{thebibliography}{35}%
\makeatletter
\providecommand \@ifxundefined [1]{%
 \@ifx{#1\undefined}
}%
\providecommand \@ifnum [1]{%
 \ifnum #1\expandafter \@firstoftwo
 \else \expandafter \@secondoftwo
 \fi
}%
\providecommand \@ifx [1]{%
 \ifx #1\expandafter \@firstoftwo
 \else \expandafter \@secondoftwo
 \fi
}%
\providecommand \natexlab [1]{#1}%
\providecommand \enquote  [1]{``#1''}%
\providecommand \bibnamefont  [1]{#1}%
\providecommand \bibfnamefont [1]{#1}%
\providecommand \citenamefont [1]{#1}%
\providecommand \href@noop [0]{\@secondoftwo}%
\providecommand \href [0]{\begingroup \@sanitize@url \@href}%
\providecommand \@href[1]{\@@startlink{#1}\@@href}%
\providecommand \@@href[1]{\endgroup#1\@@endlink}%
\providecommand \@sanitize@url [0]{\catcode `\\12\catcode `\$12\catcode
  `\&12\catcode `\#12\catcode `\^12\catcode `\_12\catcode `\%12\relax}%
\providecommand \@@startlink[1]{}%
\providecommand \@@endlink[0]{}%
\providecommand \url  [0]{\begingroup\@sanitize@url \@url }%
\providecommand \@url [1]{\endgroup\@href {#1}{\urlprefix }}%
\providecommand \urlprefix  [0]{URL }%
\providecommand \Eprint [0]{\href }%
\providecommand \doibase [0]{http://dx.doi.org/}%
\providecommand \selectlanguage [0]{\@gobble}%
\providecommand \bibinfo  [0]{\@secondoftwo}%
\providecommand \bibfield  [0]{\@secondoftwo}%
\providecommand \translation [1]{[#1]}%
\providecommand \BibitemOpen [0]{}%
\providecommand \bibitemStop [0]{}%
\providecommand \bibitemNoStop [0]{.\EOS\space}%
\providecommand \EOS [0]{\spacefactor3000\relax}%
\providecommand \BibitemShut  [1]{\csname bibitem#1\endcsname}%
\let\auto@bib@innerbib\@empty
\bibitem [{\citenamefont {Alicea}(2012)}]{Alicea2012}%
  \BibitemOpen
  \bibfield  {author} {\bibinfo {author} {\bibfnamefont {J.}~\bibnamefont
  {Alicea}},\ }\href {\doibase 10.1088/0034-4885/75/7/076501} {\bibfield
  {journal} {\bibinfo  {journal} {Rep. Prog. Phys.}\ }\textbf {\bibinfo
  {volume} {75}},\ \bibinfo {pages} {076501} (\bibinfo {year}
  {2012})}\BibitemShut {NoStop}%
\bibitem [{\citenamefont {Beenakker}(2013)}]{Beenakker2013}%
  \BibitemOpen
  \bibfield  {author} {\bibinfo {author} {\bibfnamefont {C.}~\bibnamefont
  {Beenakker}},\ }\href {\doibase 10.1146/annurev-conmatphys-030212-184337}
  {\bibfield  {journal} {\bibinfo  {journal} {Annual Review of Condensed Matter
  Physics}\ }\textbf {\bibinfo {volume} {4}},\ \bibinfo {pages} {113} (\bibinfo
  {year} {2013})}\BibitemShut {NoStop}%
\bibitem [{\citenamefont {Leijnse}\ and\ \citenamefont
  {Flensberg}(2012)}]{leijnse2012introduction}%
  \BibitemOpen
  \bibfield  {author} {\bibinfo {author} {\bibfnamefont {M.}~\bibnamefont
  {Leijnse}}\ and\ \bibinfo {author} {\bibfnamefont {K.}~\bibnamefont
  {Flensberg}},\ }\href@noop {} {\bibfield  {journal} {\bibinfo  {journal}
  {Semiconductor Science and Technology}\ }\textbf {\bibinfo {volume} {27}},\
  \bibinfo {pages} {124003} (\bibinfo {year} {2012})}\BibitemShut {NoStop}%
\bibitem [{\citenamefont {Fidkowski}\ \emph {et~al.}(2011)\citenamefont
  {Fidkowski}, \citenamefont {Lutchyn}, \citenamefont {Nayak},\ and\
  \citenamefont {Fisher}}]{Fidkowski2011}%
  \BibitemOpen
  \bibfield  {author} {\bibinfo {author} {\bibfnamefont {L.}~\bibnamefont
  {Fidkowski}}, \bibinfo {author} {\bibfnamefont {R.}~\bibnamefont {Lutchyn}},
  \bibinfo {author} {\bibfnamefont {C.}~\bibnamefont {Nayak}}, \ and\ \bibinfo
  {author} {\bibfnamefont {M.}~\bibnamefont {Fisher}},\ }\href {\doibase
  10.1103/PhysRevB.84.195436} {\bibfield  {journal} {\bibinfo  {journal} {Phys.
  Rev. B}\ }\textbf {\bibinfo {volume} {84}},\ \bibinfo {pages} {195436}
  (\bibinfo {year} {2011})}\BibitemShut {NoStop}%
\bibitem [{\citenamefont {Sau}\ \emph {et~al.}(2011)\citenamefont {Sau},
  \citenamefont {Halperin}, \citenamefont {Flensberg},\ and\ \citenamefont
  {Das~Sarma}}]{Sau2011}%
  \BibitemOpen
  \bibfield  {author} {\bibinfo {author} {\bibfnamefont {J.}~\bibnamefont
  {Sau}}, \bibinfo {author} {\bibfnamefont {B.}~\bibnamefont {Halperin}},
  \bibinfo {author} {\bibfnamefont {K.}~\bibnamefont {Flensberg}}, \ and\
  \bibinfo {author} {\bibfnamefont {S.}~\bibnamefont {Das~Sarma}},\ }\href
  {\doibase 10.1103/PhysRevB.84.144509} {\bibfield  {journal} {\bibinfo
  {journal} {Phys. Rev. B}\ }\textbf {\bibinfo {volume} {84}},\ \bibinfo
  {pages} {144509} (\bibinfo {year} {2011})}\BibitemShut {NoStop}%
\bibitem [{\citenamefont {Ruhman}\ \emph {et~al.}(2015)\citenamefont {Ruhman},
  \citenamefont {Berg},\ and\ \citenamefont {Altman}}]{Ruhman2015}%
  \BibitemOpen
  \bibfield  {author} {\bibinfo {author} {\bibfnamefont {J.}~\bibnamefont
  {Ruhman}}, \bibinfo {author} {\bibfnamefont {E.}~\bibnamefont {Berg}}, \ and\
  \bibinfo {author} {\bibfnamefont {E.}~\bibnamefont {Altman}},\ }\href
  {\doibase 10.1103/PhysRevLett.114.100401} {\bibfield  {journal} {\bibinfo
  {journal} {Phys. Rev. Lett.}\ }\textbf {\bibinfo {volume} {114}},\ \bibinfo
  {pages} {100401} (\bibinfo {year} {2015})}\BibitemShut {NoStop}%
\bibitem [{\citenamefont {Kane}\ \emph {et~al.}(2017)\citenamefont {Kane},
  \citenamefont {Stern},\ and\ \citenamefont {Halperin}}]{kane2017pairing}%
  \BibitemOpen
  \bibfield  {author} {\bibinfo {author} {\bibfnamefont {C.~L.}\ \bibnamefont
  {Kane}}, \bibinfo {author} {\bibfnamefont {A.}~\bibnamefont {Stern}}, \ and\
  \bibinfo {author} {\bibfnamefont {B.~I.}\ \bibnamefont {Halperin}},\ }\href
  {\doibase 10.1103/PhysRevX.7.031009} {\bibfield  {journal} {\bibinfo
  {journal} {Phys. Rev. X}\ }\textbf {\bibinfo {volume} {7}},\ \bibinfo {pages}
  {031009} (\bibinfo {year} {2017})}\BibitemShut {NoStop}%
\bibitem [{\citenamefont {Starykh}\ \emph {et~al.}(2000)\citenamefont
  {Starykh}, \citenamefont {Maslov}, \citenamefont {H{\"a}usler}, \citenamefont
  {Glazman},\ and\ \citenamefont {Glazman}}]{Starykh2000}%
  \BibitemOpen
  \bibfield  {author} {\bibinfo {author} {\bibfnamefont {O.~A.}\ \bibnamefont
  {Starykh}}, \bibinfo {author} {\bibfnamefont {D.~L.}\ \bibnamefont {Maslov}},
  \bibinfo {author} {\bibfnamefont {W.}~\bibnamefont {H{\"a}usler}}, \bibinfo
  {author} {\bibfnamefont {L.~I.}\ \bibnamefont {Glazman}}, \ and\ \bibinfo
  {author} {\bibnamefont {Glazman}},\ }in\ \href@noop {} {\emph {\bibinfo
  {booktitle} {Low-Dimensional Systems}}},\ \bibinfo {editor} {edited by\
  \bibinfo {editor} {\bibfnamefont {T.}~\bibnamefont {Brandes}}}\ (\bibinfo
  {publisher} {Springer Berlin Heidelberg},\ \bibinfo {address} {Berlin,
  Heidelberg},\ \bibinfo {year} {2000})\ pp.\ \bibinfo {pages}
  {37--78}\BibitemShut {NoStop}%
\bibitem [{\citenamefont {Sun}\ \emph {et~al.}(2007)\citenamefont {Sun},
  \citenamefont {Gangadharaiah},\ and\ \citenamefont {Starykh}}]{Starykh2007}%
  \BibitemOpen
  \bibfield  {author} {\bibinfo {author} {\bibfnamefont {J.}~\bibnamefont
  {Sun}}, \bibinfo {author} {\bibfnamefont {S.}~\bibnamefont {Gangadharaiah}},
  \ and\ \bibinfo {author} {\bibfnamefont {O.~A.}\ \bibnamefont {Starykh}},\
  }\href {\doibase 10.1103/PhysRevLett.98.126408} {\bibfield  {journal}
  {\bibinfo  {journal} {Phys. Rev. Lett.}\ }\textbf {\bibinfo {volume} {98}},\
  \bibinfo {pages} {126408} (\bibinfo {year} {2007})}\BibitemShut {NoStop}%
\bibitem [{\citenamefont {Ruhman}\ \emph {et~al.}(2012)\citenamefont {Ruhman},
  \citenamefont {Dalla~Torre}, \citenamefont {Huber},\ and\ \citenamefont
  {Altman}}]{Ruhman2012}%
  \BibitemOpen
  \bibfield  {author} {\bibinfo {author} {\bibfnamefont {J.}~\bibnamefont
  {Ruhman}}, \bibinfo {author} {\bibfnamefont {E.~G.}\ \bibnamefont
  {Dalla~Torre}}, \bibinfo {author} {\bibfnamefont {S.~D.}\ \bibnamefont
  {Huber}}, \ and\ \bibinfo {author} {\bibfnamefont {E.}~\bibnamefont
  {Altman}},\ }\href {\doibase 10.1103/PhysRevB.85.125121} {\bibfield
  {journal} {\bibinfo  {journal} {Phys. Rev. B}\ }\textbf {\bibinfo {volume}
  {85}},\ \bibinfo {pages} {125121} (\bibinfo {year} {2012})}\BibitemShut
  {NoStop}%
\bibitem [{\citenamefont {Kraus}\ \emph {et~al.}(2013)\citenamefont {Kraus},
  \citenamefont {Dalmonte}, \citenamefont {Baranov}, \citenamefont
  {L\"auchli},\ and\ \citenamefont {Zoller}}]{Zoller2013}%
  \BibitemOpen
  \bibfield  {author} {\bibinfo {author} {\bibfnamefont {C.~V.}\ \bibnamefont
  {Kraus}}, \bibinfo {author} {\bibfnamefont {M.}~\bibnamefont {Dalmonte}},
  \bibinfo {author} {\bibfnamefont {M.~A.}\ \bibnamefont {Baranov}}, \bibinfo
  {author} {\bibfnamefont {A.~M.}\ \bibnamefont {L\"auchli}}, \ and\ \bibinfo
  {author} {\bibfnamefont {P.}~\bibnamefont {Zoller}},\ }\href {\doibase
  10.1103/PhysRevLett.111.173004} {\bibfield  {journal} {\bibinfo  {journal}
  {Phys. Rev. Lett.}\ }\textbf {\bibinfo {volume} {111}},\ \bibinfo {pages}
  {173004} (\bibinfo {year} {2013})}\BibitemShut {NoStop}%
\bibitem [{\citenamefont {Keselman}\ and\ \citenamefont
  {Berg}(2015)}]{Keselman2015}%
  \BibitemOpen
  \bibfield  {author} {\bibinfo {author} {\bibfnamefont {A.}~\bibnamefont
  {Keselman}}\ and\ \bibinfo {author} {\bibfnamefont {E.}~\bibnamefont
  {Berg}},\ }\href {\doibase 10.1103/PhysRevB.91.235309} {\bibfield  {journal}
  {\bibinfo  {journal} {Phys. Rev. B}\ }\textbf {\bibinfo {volume} {91}},\
  \bibinfo {pages} {235309} (\bibinfo {year} {2015})}\BibitemShut {NoStop}%
\bibitem [{\citenamefont {Iemini}\ \emph {et~al.}(2015)\citenamefont {Iemini},
  \citenamefont {Mazza}, \citenamefont {Rossini}, \citenamefont {Fazio},\ and\
  \citenamefont {Diehl}}]{Diehl2015}%
  \BibitemOpen
  \bibfield  {author} {\bibinfo {author} {\bibfnamefont {F.}~\bibnamefont
  {Iemini}}, \bibinfo {author} {\bibfnamefont {L.}~\bibnamefont {Mazza}},
  \bibinfo {author} {\bibfnamefont {D.}~\bibnamefont {Rossini}}, \bibinfo
  {author} {\bibfnamefont {R.}~\bibnamefont {Fazio}}, \ and\ \bibinfo {author}
  {\bibfnamefont {S.}~\bibnamefont {Diehl}},\ }\href {\doibase
  10.1103/PhysRevLett.115.156402} {\bibfield  {journal} {\bibinfo  {journal}
  {Phys. Rev. Lett.}\ }\textbf {\bibinfo {volume} {115}},\ \bibinfo {pages}
  {156402} (\bibinfo {year} {2015})}\BibitemShut {NoStop}%
\bibitem [{\citenamefont {Chen}\ \emph {et~al.}(2017)\citenamefont {Chen},
  \citenamefont {Yan}, \citenamefont {Ting}, \citenamefont {Chen},\ and\
  \citenamefont {Burnell}}]{chen2017flux}%
  \BibitemOpen
  \bibfield  {author} {\bibinfo {author} {\bibfnamefont {C.}~\bibnamefont
  {Chen}}, \bibinfo {author} {\bibfnamefont {W.}~\bibnamefont {Yan}}, \bibinfo
  {author} {\bibfnamefont {C.}~\bibnamefont {Ting}}, \bibinfo {author}
  {\bibfnamefont {Y.}~\bibnamefont {Chen}}, \ and\ \bibinfo {author}
  {\bibfnamefont {F.}~\bibnamefont {Burnell}},\ }\href@noop {} {\bibfield
  {journal} {\bibinfo  {journal} {arXiv preprint arXiv:1701.01794}\ } (\bibinfo
  {year} {2017})}\BibitemShut {NoStop}%
\bibitem [{\citenamefont {Kainaris}\ \emph
  {et~al.}(2017{\natexlab{a}})\citenamefont {Kainaris}, \citenamefont {Santos},
  \citenamefont {Gutman},\ and\ \citenamefont
  {Carr}}]{kainaris2017interaction}%
  \BibitemOpen
  \bibfield  {author} {\bibinfo {author} {\bibfnamefont {N.}~\bibnamefont
  {Kainaris}}, \bibinfo {author} {\bibfnamefont {R.~A.}\ \bibnamefont
  {Santos}}, \bibinfo {author} {\bibfnamefont {D.~B.}\ \bibnamefont {Gutman}},
  \ and\ \bibinfo {author} {\bibfnamefont {S.~T.}\ \bibnamefont {Carr}},\
  }\href {\doibase 10.1002/prop.201600054} {\bibfield  {journal} {\bibinfo
  {journal} {Fortschritte der Physik}\ }\textbf {\bibinfo {volume} {65}},\
  \bibinfo {pages} {1600054} (\bibinfo {year} {2017}{\natexlab{a}})},\ \bibinfo
  {note} {1600054}\BibitemShut {NoStop}%
\bibitem [{\citenamefont {Kainaris}\ \emph
  {et~al.}(2017{\natexlab{b}})\citenamefont {Kainaris}, \citenamefont {Carr},\
  and\ \citenamefont {Mirlin}}]{kainaris2017transmission}%
  \BibitemOpen
  \bibfield  {author} {\bibinfo {author} {\bibfnamefont {N.}~\bibnamefont
  {Kainaris}}, \bibinfo {author} {\bibfnamefont {S.~T.}\ \bibnamefont {Carr}},
  \ and\ \bibinfo {author} {\bibfnamefont {A.~D.}\ \bibnamefont {Mirlin}},\
  }\href@noop {} {\bibfield  {journal} {\bibinfo  {journal} {arXiv preprint
  arXiv:1709.08965}\ } (\bibinfo {year} {2017}{\natexlab{b}})}\BibitemShut
  {NoStop}%
\bibitem [{\citenamefont {Scaffidi}\ \emph {et~al.}(2017)\citenamefont
  {Scaffidi}, \citenamefont {Parker},\ and\ \citenamefont
  {Vasseur}}]{Scaffidi2017}%
  \BibitemOpen
  \bibfield  {author} {\bibinfo {author} {\bibfnamefont {T.}~\bibnamefont
  {Scaffidi}}, \bibinfo {author} {\bibfnamefont {D.~E.}\ \bibnamefont
  {Parker}}, \ and\ \bibinfo {author} {\bibfnamefont {R.}~\bibnamefont
  {Vasseur}},\ }\href {\doibase 10.1103/PhysRevX.7.041048} {\bibfield
  {journal} {\bibinfo  {journal} {Phys. Rev. X}\ }\textbf {\bibinfo {volume}
  {7}},\ \bibinfo {pages} {041048} (\bibinfo {year} {2017})}\BibitemShut
  {NoStop}%
\bibitem [{\citenamefont {Jiang}\ \emph {et~al.}(2017)\citenamefont {Jiang},
  \citenamefont {Li}, \citenamefont {Seidel},\ and\ \citenamefont
  {Lee}}]{jiang2017symmetry}%
  \BibitemOpen
  \bibfield  {author} {\bibinfo {author} {\bibfnamefont {H.-C.}\ \bibnamefont
  {Jiang}}, \bibinfo {author} {\bibfnamefont {Z.-X.}\ \bibnamefont {Li}},
  \bibinfo {author} {\bibfnamefont {A.}~\bibnamefont {Seidel}}, \ and\ \bibinfo
  {author} {\bibfnamefont {D.-H.}\ \bibnamefont {Lee}},\ }\href@noop {}
  {\bibfield  {journal} {\bibinfo  {journal} {arXiv preprint arXiv:1704.02997}\
  } (\bibinfo {year} {2017})}\BibitemShut {NoStop}%
\bibitem [{\citenamefont {White}(1992)}]{White1992PRL}%
  \BibitemOpen
  \bibfield  {author} {\bibinfo {author} {\bibfnamefont {S.~R.}\ \bibnamefont
  {White}},\ }\href {\doibase 10.1103/PhysRevLett.69.2863} {\bibfield
  {journal} {\bibinfo  {journal} {Phys. Rev. Lett.}\ }\textbf {\bibinfo
  {volume} {69}},\ \bibinfo {pages} {2863} (\bibinfo {year}
  {1992})}\BibitemShut {NoStop}%
\bibitem [{\citenamefont {White}(1993)}]{White1992PRB}%
  \BibitemOpen
  \bibfield  {author} {\bibinfo {author} {\bibfnamefont {S.~R.}\ \bibnamefont
  {White}},\ }\href {\doibase 10.1103/PhysRevB.48.10345} {\bibfield  {journal}
  {\bibinfo  {journal} {Phys. Rev. B}\ }\textbf {\bibinfo {volume} {48}},\
  \bibinfo {pages} {10345} (\bibinfo {year} {1993})}\BibitemShut {NoStop}%
\bibitem [{\citenamefont {Yamanaka}\ \emph {et~al.}(1997)\citenamefont
  {Yamanaka}, \citenamefont {Oshikawa},\ and\ \citenamefont
  {Affleck}}]{OshikawaAffleck}%
  \BibitemOpen
  \bibfield  {author} {\bibinfo {author} {\bibfnamefont {M.}~\bibnamefont
  {Yamanaka}}, \bibinfo {author} {\bibfnamefont {M.}~\bibnamefont {Oshikawa}},
  \ and\ \bibinfo {author} {\bibfnamefont {I.}~\bibnamefont {Affleck}},\ }\href
  {\doibase 10.1103/PhysRevLett.79.1110} {\bibfield  {journal} {\bibinfo
  {journal} {Phys. Rev. Lett.}\ }\textbf {\bibinfo {volume} {79}},\ \bibinfo
  {pages} {1110} (\bibinfo {year} {1997})}\BibitemShut {NoStop}%
\bibitem [{\citenamefont {Ruhman}\ and\ \citenamefont
  {Altman}(2017)}]{ruhman2017topological}%
  \BibitemOpen
  \bibfield  {author} {\bibinfo {author} {\bibfnamefont {J.}~\bibnamefont
  {Ruhman}}\ and\ \bibinfo {author} {\bibfnamefont {E.}~\bibnamefont
  {Altman}},\ }\href {\doibase 10.1103/PhysRevB.96.085133} {\bibfield
  {journal} {\bibinfo  {journal} {Phys. Rev. B}\ }\textbf {\bibinfo {volume}
  {96}},\ \bibinfo {pages} {085133} (\bibinfo {year} {2017})}\BibitemShut
  {NoStop}%
\bibitem [{\citenamefont {Altland}\ and\ \citenamefont
  {Zirnbauer}(1997)}]{AltlandZirnbauer1997}%
  \BibitemOpen
  \bibfield  {author} {\bibinfo {author} {\bibfnamefont {A.}~\bibnamefont
  {Altland}}\ and\ \bibinfo {author} {\bibfnamefont {M.~R.}\ \bibnamefont
  {Zirnbauer}},\ }\href {\doibase 10.1103/PhysRevB.55.1142} {\bibfield
  {journal} {\bibinfo  {journal} {Physical Review B}\ }\textbf {\bibinfo
  {volume} {55}},\ \bibinfo {pages} {1142} (\bibinfo {year}
  {1997})}\BibitemShut {NoStop}%
\bibitem [{\citenamefont {Kitaev}(2001)}]{Kitaev2001}%
  \BibitemOpen
  \bibfield  {author} {\bibinfo {author} {\bibfnamefont {A.~Y.}\ \bibnamefont
  {Kitaev}},\ }\href {\doibase 10.1070/1063-7869/44/10S/S29} {\bibfield
  {journal} {\bibinfo  {journal} {Physics-Uspekhi}\ }\textbf {\bibinfo {volume}
  {44}},\ \bibinfo {pages} {131} (\bibinfo {year} {2001})}\BibitemShut
  {NoStop}%
\bibitem [{\citenamefont {Azaria}(2017)}]{AzariaLuttinger}%
  \BibitemOpen
  \bibfield  {author} {\bibinfo {author} {\bibfnamefont {P.}~\bibnamefont
  {Azaria}},\ }\href@noop {} {\bibfield  {journal} {\bibinfo  {journal} {Phys.
  Rev. B}\ }\textbf {\bibinfo {volume} {95}},\ \bibinfo {pages} {125106}
  (\bibinfo {year} {2017})}\BibitemShut {NoStop}%
\bibitem [{\citenamefont {Itoi}\ and\ \citenamefont {Kato}(1997)}]{Itoi1997}%
  \BibitemOpen
  \bibfield  {author} {\bibinfo {author} {\bibfnamefont {C.}~\bibnamefont
  {Itoi}}\ and\ \bibinfo {author} {\bibfnamefont {M.-H.}\ \bibnamefont
  {Kato}},\ }\href {\doibase 10.1103/PhysRevB.55.8295} {\bibfield  {journal}
  {\bibinfo  {journal} {Phys. Rev. B}\ }\textbf {\bibinfo {volume} {55}},\
  \bibinfo {pages} {8295} (\bibinfo {year} {1997})}\BibitemShut {NoStop}%
\bibitem [{\citenamefont {Lecheminant}\ \emph {et~al.}(2008)\citenamefont
  {Lecheminant}, \citenamefont {Azaria}, \citenamefont {Boulat}, \citenamefont
  {Capponi}, \citenamefont {Roux},\ and\ \citenamefont
  {White}}]{LecheminantTrions}%
  \BibitemOpen
  \bibfield  {author} {\bibinfo {author} {\bibfnamefont {P.}~\bibnamefont
  {Lecheminant}}, \bibinfo {author} {\bibfnamefont {P.}~\bibnamefont {Azaria}},
  \bibinfo {author} {\bibfnamefont {E.}~\bibnamefont {Boulat}}, \bibinfo
  {author} {\bibfnamefont {S.}~\bibnamefont {Capponi}}, \bibinfo {author}
  {\bibfnamefont {G.}~\bibnamefont {Roux}}, \ and\ \bibinfo {author}
  {\bibfnamefont {S.}~\bibnamefont {White}},\ }\href@noop {} {\bibfield
  {journal} {\bibinfo  {journal} {I. J. Mod. Phys. E}\ }\textbf {\bibinfo
  {volume} {17}},\ \bibinfo {pages} {2110} (\bibinfo {year}
  {2008})}\BibitemShut {NoStop}%
\bibitem [{\citenamefont {Boulat}\ \emph {et~al.}(2009)\citenamefont {Boulat},
  \citenamefont {Azaria},\ and\ \citenamefont {Lecheminant}}]{BoulatDuality}%
  \BibitemOpen
  \bibfield  {author} {\bibinfo {author} {\bibfnamefont {E.}~\bibnamefont
  {Boulat}}, \bibinfo {author} {\bibfnamefont {P.}~\bibnamefont {Azaria}}, \
  and\ \bibinfo {author} {\bibfnamefont {P.}~\bibnamefont {Lecheminant}},\
  }\href@noop {} {\bibfield  {journal} {\bibinfo  {journal} {Nucl. Phys. B}\
  }\textbf {\bibinfo {volume} {822}},\ \bibinfo {pages} {367} (\bibinfo {year}
  {2009})}\BibitemShut {NoStop}%
\bibitem [{\citenamefont {Andrei}\ and\ \citenamefont
  {Destri}(1984)}]{Andrei1984}%
  \BibitemOpen
  \bibfield  {author} {\bibinfo {author} {\bibfnamefont {N.}~\bibnamefont
  {Andrei}}\ and\ \bibinfo {author} {\bibfnamefont {C.}~\bibnamefont
  {Destri}},\ }\href {\doibase http://dx.doi.org/10.1016/0550-3213(84)90514-5}
  {\bibfield  {journal} {\bibinfo  {journal} {Nuclear Physics B}\ }\textbf
  {\bibinfo {volume} {231}},\ \bibinfo {pages} {445 } (\bibinfo {year}
  {1984})}\BibitemShut {NoStop}%
\bibitem [{\citenamefont {Assaraf}\ \emph {et~al.}(1999)\citenamefont
  {Assaraf}, \citenamefont {Azaria}, \citenamefont {Caffarel},\ and\
  \citenamefont {Lecheminant}}]{RolandAzaria1999}%
  \BibitemOpen
  \bibfield  {author} {\bibinfo {author} {\bibfnamefont {R.}~\bibnamefont
  {Assaraf}}, \bibinfo {author} {\bibfnamefont {P.}~\bibnamefont {Azaria}},
  \bibinfo {author} {\bibfnamefont {M.}~\bibnamefont {Caffarel}}, \ and\
  \bibinfo {author} {\bibfnamefont {P.}~\bibnamefont {Lecheminant}},\ }\href
  {\doibase 10.1103/PhysRevB.60.2299} {\bibfield  {journal} {\bibinfo
  {journal} {Phys. Rev. B}\ }\textbf {\bibinfo {volume} {60}},\ \bibinfo
  {pages} {2299} (\bibinfo {year} {1999})}\BibitemShut {NoStop}%
\bibitem [{\citenamefont {Schollw\"ock}\ \emph {et~al.}(1996)\citenamefont
  {Schollw\"ock}, \citenamefont {Jolic\oe{}ur},\ and\ \citenamefont
  {Garel}}]{Schollwock1996}%
  \BibitemOpen
  \bibfield  {author} {\bibinfo {author} {\bibfnamefont {U.}~\bibnamefont
  {Schollw\"ock}}, \bibinfo {author} {\bibfnamefont {T.}~\bibnamefont
  {Jolic\oe{}ur}}, \ and\ \bibinfo {author} {\bibfnamefont {T.}~\bibnamefont
  {Garel}},\ }\href {\doibase 10.1103/PhysRevB.53.3304} {\bibfield  {journal}
  {\bibinfo  {journal} {Phys. Rev. B}\ }\textbf {\bibinfo {volume} {53}},\
  \bibinfo {pages} {3304} (\bibinfo {year} {1996})}\BibitemShut {NoStop}%
\bibitem [{\citenamefont {Tu}\ \emph {et~al.}(2008)\citenamefont {Tu},
  \citenamefont {Zhang},\ and\ \citenamefont {Xiang}}]{HongHao2008}%
  \BibitemOpen
  \bibfield  {author} {\bibinfo {author} {\bibfnamefont {H.-H.}\ \bibnamefont
  {Tu}}, \bibinfo {author} {\bibfnamefont {G.-M.}\ \bibnamefont {Zhang}}, \
  and\ \bibinfo {author} {\bibfnamefont {T.}~\bibnamefont {Xiang}},\ }\href
  {\doibase 10.1103/PhysRevB.78.094404} {\bibfield  {journal} {\bibinfo
  {journal} {Phys. Rev. B}\ }\textbf {\bibinfo {volume} {78}},\ \bibinfo
  {pages} {094404} (\bibinfo {year} {2008})}\BibitemShut {NoStop}%
\bibitem [{\citenamefont {Affleck}\ \emph {et~al.}(1987)\citenamefont
  {Affleck}, \citenamefont {Kennedy}, \citenamefont {Lieb},\ and\ \citenamefont
  {Tasaki}}]{AKLT1987}%
  \BibitemOpen
  \bibfield  {author} {\bibinfo {author} {\bibfnamefont {I.}~\bibnamefont
  {Affleck}}, \bibinfo {author} {\bibfnamefont {T.}~\bibnamefont {Kennedy}},
  \bibinfo {author} {\bibfnamefont {E.~H.}\ \bibnamefont {Lieb}}, \ and\
  \bibinfo {author} {\bibfnamefont {H.}~\bibnamefont {Tasaki}},\ }\href
  {\doibase 10.1103/PhysRevLett.59.799} {\bibfield  {journal} {\bibinfo
  {journal} {Phys. Rev. Lett.}\ }\textbf {\bibinfo {volume} {59}},\ \bibinfo
  {pages} {799} (\bibinfo {year} {1987})}\BibitemShut {NoStop}%
\bibitem [{ITe()}]{ITensor}%
  \BibitemOpen
  \href@noop {} {}\bibinfo {note} {Calculations were performed using the
  ITensor Library, \href{http://itensor.org/}{http://itensor.org/}}\BibitemShut
  {NoStop}%
\bibitem [{\citenamefont {Chepiga}\ \emph {et~al.}(2016)\citenamefont
  {Chepiga}, \citenamefont {Affleck},\ and\ \citenamefont
  {Mila}}]{Chepiga2016}%
  \BibitemOpen
  \bibfield  {author} {\bibinfo {author} {\bibfnamefont {N.}~\bibnamefont
  {Chepiga}}, \bibinfo {author} {\bibfnamefont {I.}~\bibnamefont {Affleck}}, \
  and\ \bibinfo {author} {\bibfnamefont {F.}~\bibnamefont {Mila}},\ }\href
  {\doibase 10.1103/PhysRevB.93.241108} {\bibfield  {journal} {\bibinfo
  {journal} {Phys. Rev. B}\ }\textbf {\bibinfo {volume} {93}},\ \bibinfo
  {pages} {241108} (\bibinfo {year} {2016})}\BibitemShut {NoStop}%
\end{thebibliography}%

\appendix

\section{Additional numerical results\label{appx:AdditionalDMRGResults}}

\subsection{Further analysis of the gapless Haldane phase}

In this appendix, we consider the spin and the charge sectors in the
gapless Haldane phase, showing that a finite gap opens in the former
while the latter remains gapless.

\begin{figure}
\centering
\subfloat[\label{fig:SpinGap}]{\includegraphics[width=0.7\columnwidth]{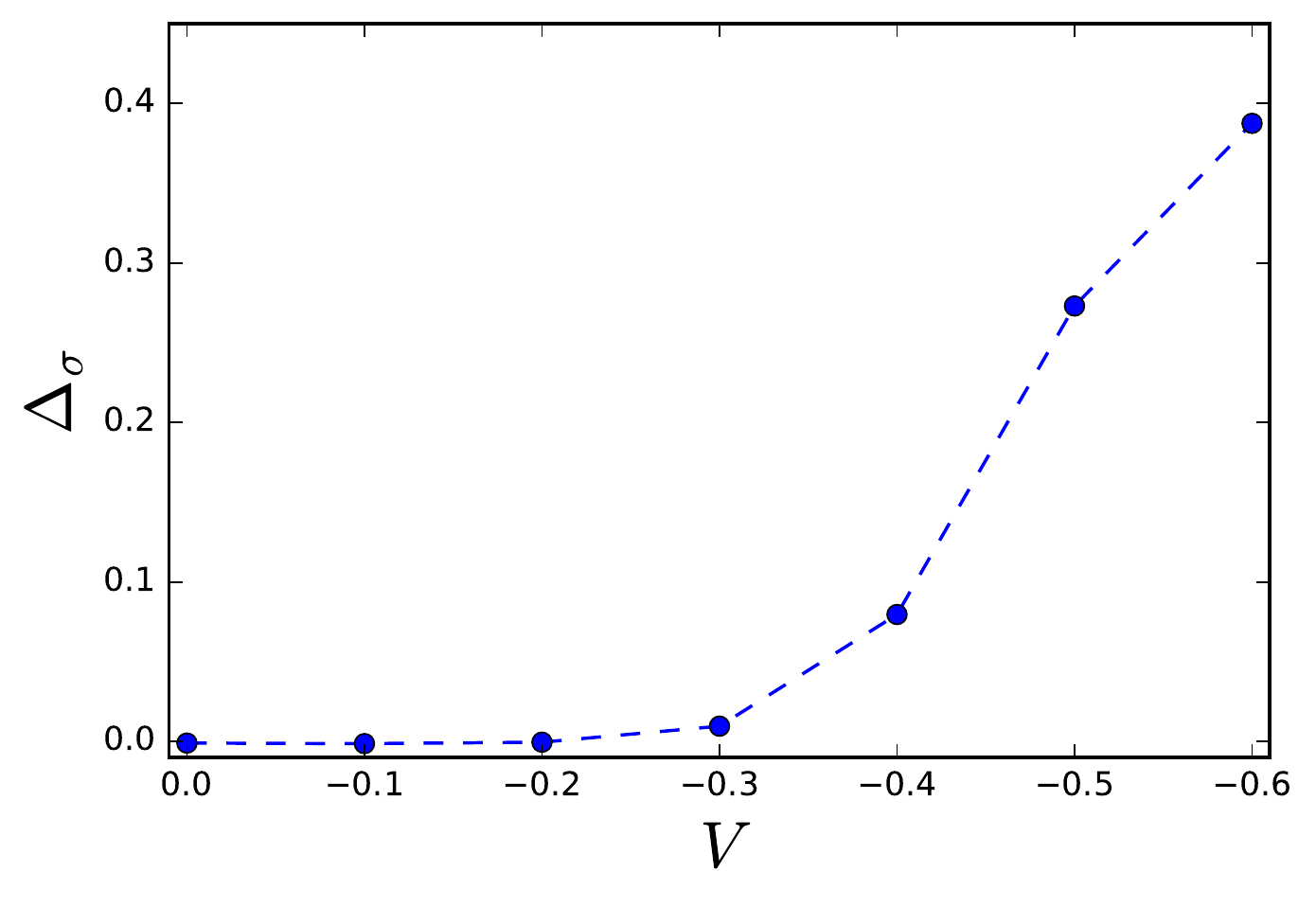}}

\subfloat[\label{fig:LuttParam}]{\includegraphics[width=0.7\columnwidth]{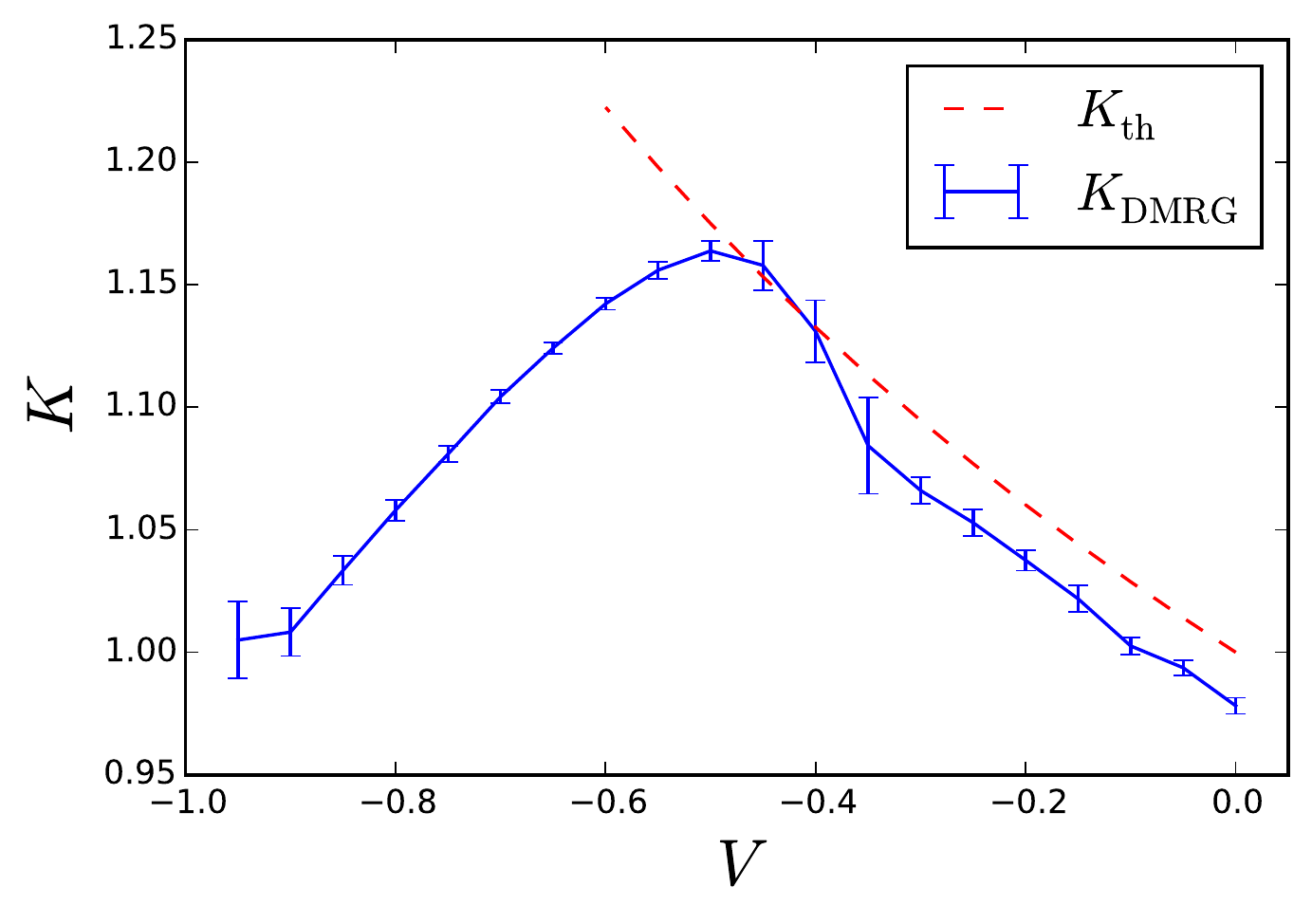}}

\caption{(a) The spin gap, $\Delta_{\sigma}$ (defined in Eq.~\eqref{eq:SpinGap})
and (b) the charge Luttinger parameter, $K$, for $U=0$ as function
of $V$. (a) For $V\lesssim-0.3$ a finite spin gap is observed. (b)
The Luttinger parameter in the charge sector obtained numerically
(blue solid line), compared to the analytical value obtained using
weak coupling analysis (red dashed line).}
\end{figure}

\subsubsection{Spin gap\label{appx:SpinGap}}

Below, we calculate the spin gap in the bulk as $\left|V\right|$
is increased. To this end, we calculate the energy gap to $S^{z}=+2$
excitations, as we expect the gap to $S^{z}=+1$ excitations to vanish
for a system with open boundary conditions due to the spin-$1/2$
end modes. More specifically, we calculate 
\be
\Delta_{\sigma}=\lim_{N_x\to\infty}\left[E_{\left(n+1,n,n-1\right)}-E_{\left(n,n,n\right)}\right]_{n=\frac{N_x}{3}},\label{eq:SpinGap}
\ee
where we denote by $E_{\left(n_{+1},n_{0},n_{-1}\right)}$ the ground
state energy of a system with $n_{+1,0,-1}$ particles of flavor
$+1$, $0$ and $-1$ respectively. The spin gap, $\Delta_{\sigma}$,
for $U=0$ as function of $V$ is shown in Fig. \ref{fig:SpinGap}.
It can be seen that, indeed, a finite spin gap opens as $\left|V\right|$
is increased. 

\subsubsection{Charge Luttinger parameter\label{appx:LuttParam}}

To verify that the charge sector is indeed gapless in the region we
identify as the gapless Haldane phase, we calculate the charge Luttinger
parameter numerically. 

To this end, we first calculate the energy of the first excited state
in the $S_{z}=1$ sector as function of system size. Assuming the
spin sector is gapped, the energy of the first excited state is given
by $\pi v_{c}/L$, allowing us to extract the value of the
charge velocity $v_c$. (In practice, since the spin gap for small values
of $V$ is small, for small systems sizes the spin sector will appear
gapless with the corresponding velocity $v_{\sigma}$, and the first excited state will be given by ${\rm min}\left(\pi v_{c}/L,\pi v_{\sigma}/L\right)$,
allowing us to obtain only a lower bound for $v_{c})$. Next, we calculate
the energy gap to adding a single particle of each species, as function
of system size, i.e. 
\begin{multline}
E_{3n+3}+E_{3n-3}-2E_{3n}=\\
\frac{1}{L}\frac{\pi v_{c}}{2K}\left[\left(3n+3\right)^{2}+\left(3n-3\right)^{2}-2\left(3n\right)^{2}\right] =\frac{1}{L}\frac{9\pi v_{c}}{NK} 
\end{multline}
Here $E_{3n}=E_{\left(n,n,n\right)}$, with the total number of particles $3n$ equal to the number of lattice sites $N_x$, and $N=3$ is the number
of flavors. (We add a single particle of each species to avoid excitations
of the spin sector at small values of $V$, when the spin gap is small.)
The extracted charge Luttinger parameter is plotted in Fig. \ref{fig:LuttParam}. 

The analytical value of $K$ obtained from weak coupling analysis
(see Sec. \ref{sec:Field-theory-analysis} in main text), given by
$K=\left(1+\bar{V} / \pi v_{F}\right)^{-1/2}$ for $U=0$,
is shown on the same plot for comparison. 
The critical value of the charge Luttinger parameter $K$, at which a charge gap is expected to open, is equal to $2/3$. 
Although we observe the general trend of the decrease in $K$ with increasing $\left|V\right|$ numerically (for $V\lesssim-0.5$), we could not obtain the value of $K$ close to the charge gap opening point with good enough precision to validate this.

\subsection{Determining the phase boundary of the gapless Haldane phase\label{appx:ChargePB}}

To obtain the phase boundary between the gapless and the fully gapped
Haldane phases (i.e. the Mott transition point), we calculate the
gap to adding a single particle as function of $U$ for each value of $V$, and obtain the critical value of $U$,
$U_{c}\left(V\right)$, for which this gap becomes finite. 

More specifically, we calculate $E_{3n}^{\rm SP}$ (see Eq.~\eqref{eq:E_SP}
in main text), for different system sizes up to $N_x=48$ sites (where we take the total number of particles $3n$ to be equal the number of lattices sites $N_x$), and
extrapolate it to the infinite system size limit, denoting $E^{\rm SP}=\lim_{N_x\to\infty}E_{3n}^{\rm SP}$. 
(The single particle gap is calculated in the $S^{z}=0$ sector,
i.e. we calculate $E_{\left(n,n+1,n\right)}+E_{\left(n,n-1,n\right)}-2E_{\left(n,n,n\right)}$.) 

\begin{figure}
\centering
\subfloat[]{\includegraphics[width=0.7\columnwidth]{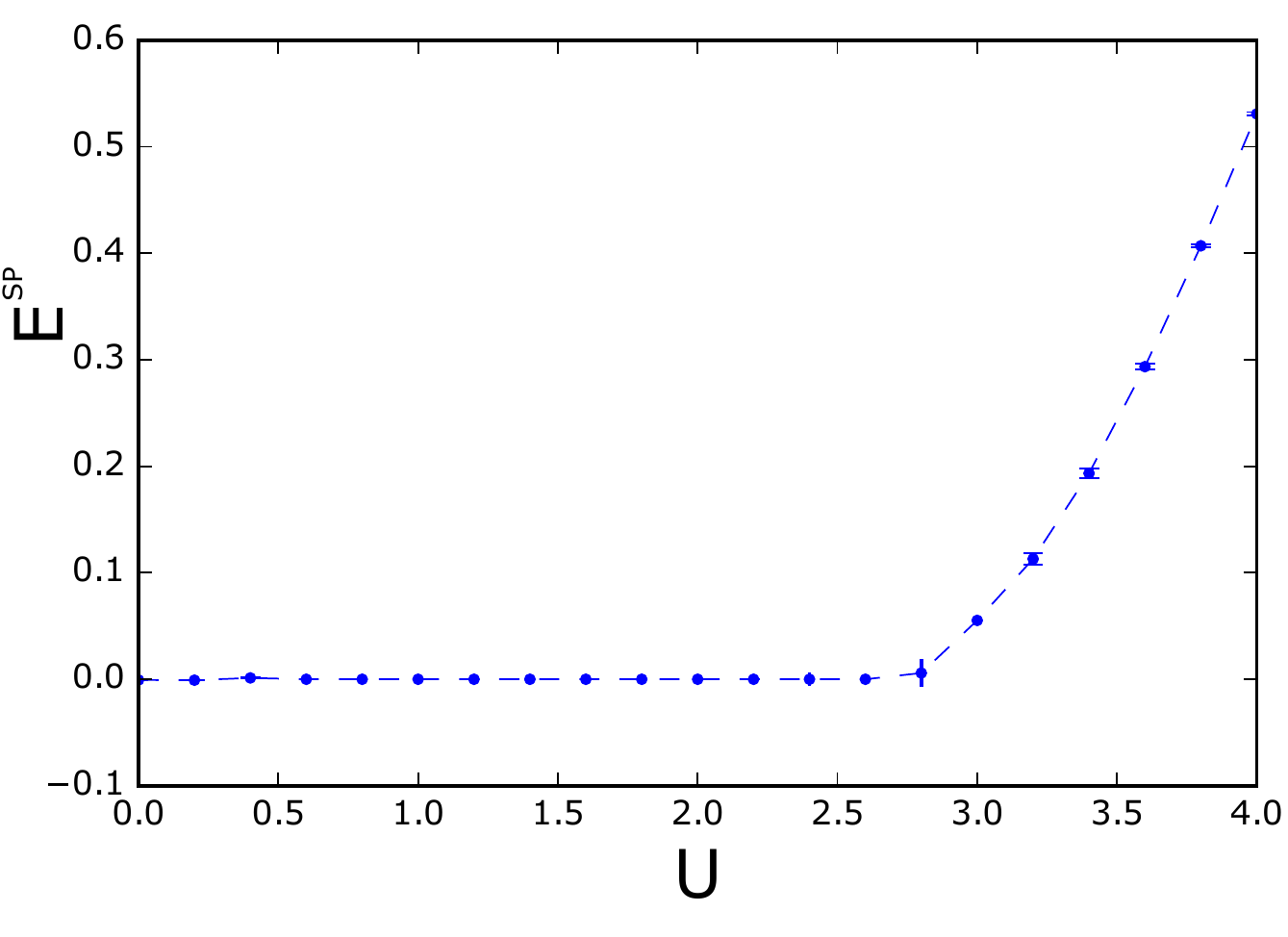}}

\subfloat[]{\includegraphics[width=0.7\columnwidth]{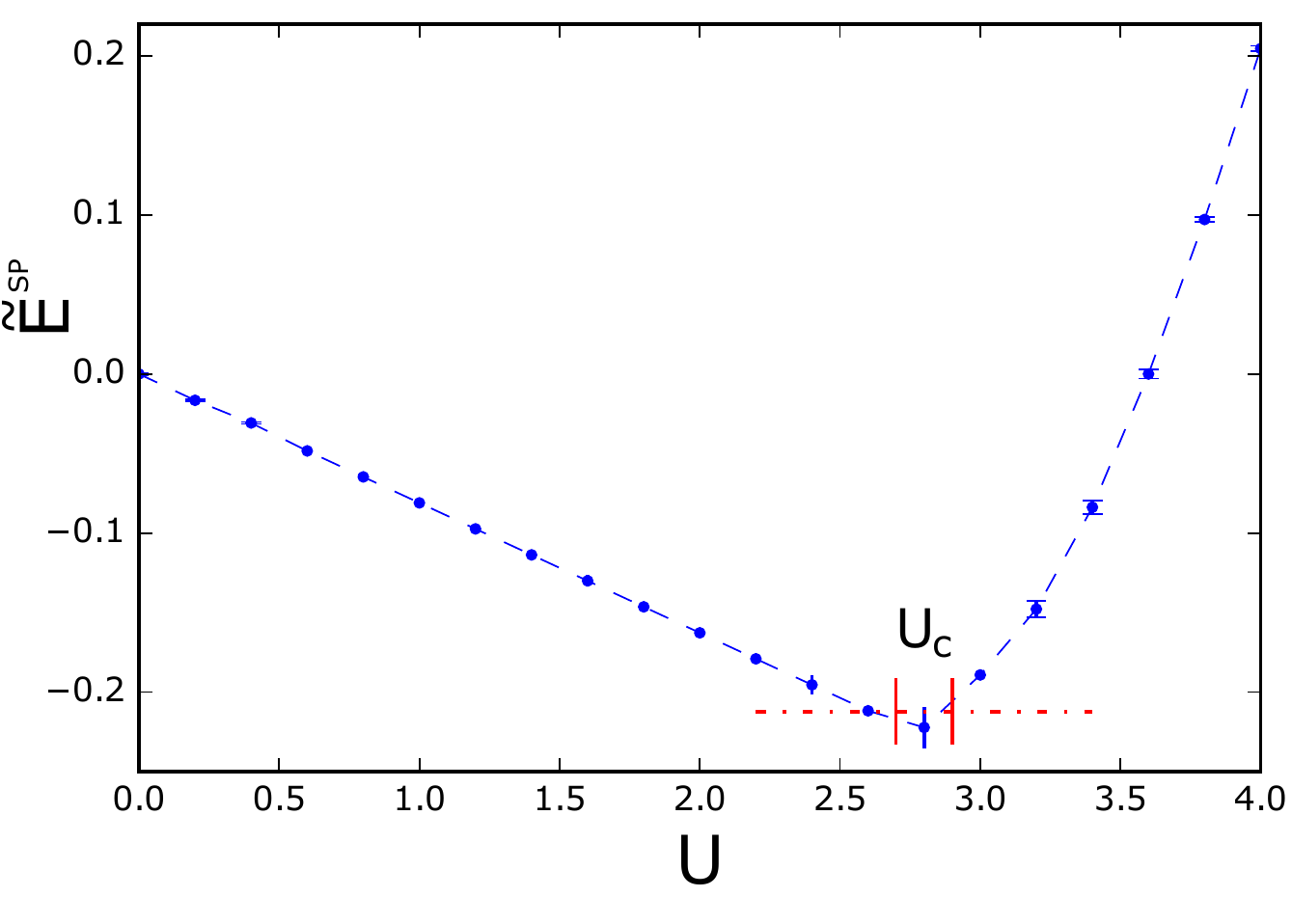}}

\caption{The (a) single particle gap, $E^{\rm SP}\left(U\right)$, and (b) the
transformed function $\tilde{E}^{\rm SP}\left(U\right)$ (see Eq.~\eqref{eq:ESP_transformed})
for $V=-0.2.$ The transformed function exhibits a minimum at the
point where the single particle gap becomes finite. The error in $U_{c}$
is estimated from the error in $E^{\rm SP}\left(U_{c}\right)$ as depicted
in the figure. \label{fig:ChargePB}}
\end{figure}

Numerically, it is difficult to obtain the charge gap opening point
directly from the function $E^{\rm SP}\left(U\right)$. Instead, we consider
the function 
\begin{multline}
\tilde{E}^{\rm SP}\left(U\right)=E^{\rm SP}\left(U\right)-E^{\rm SP}\left(U=0\right)- \\ 
\frac{E^{\rm SP}\left(U=U_{0}\right)-E^{\rm SP}\left(U=0\right)}{U_{0}}U,\label{eq:ESP_transformed}
\end{multline}
where $U_{0}>U_{c}$. (The value of $U_{0}$ is somewhat arbitrary
and we choose it such that $E^{\rm SP}\left(U=U_{0}\right)>0.2$.) This
function has an extremum point at $U_{c}$, making it easier to identify
numerically (see Fig. \ref{fig:ChargePB}).

\subsection{Dimerization transition\label{appx:DimerPB}}

\begin{figure}
\centering
\subfloat[\label{fig:dimer_finitesize}]{\includegraphics[width=0.7\columnwidth]{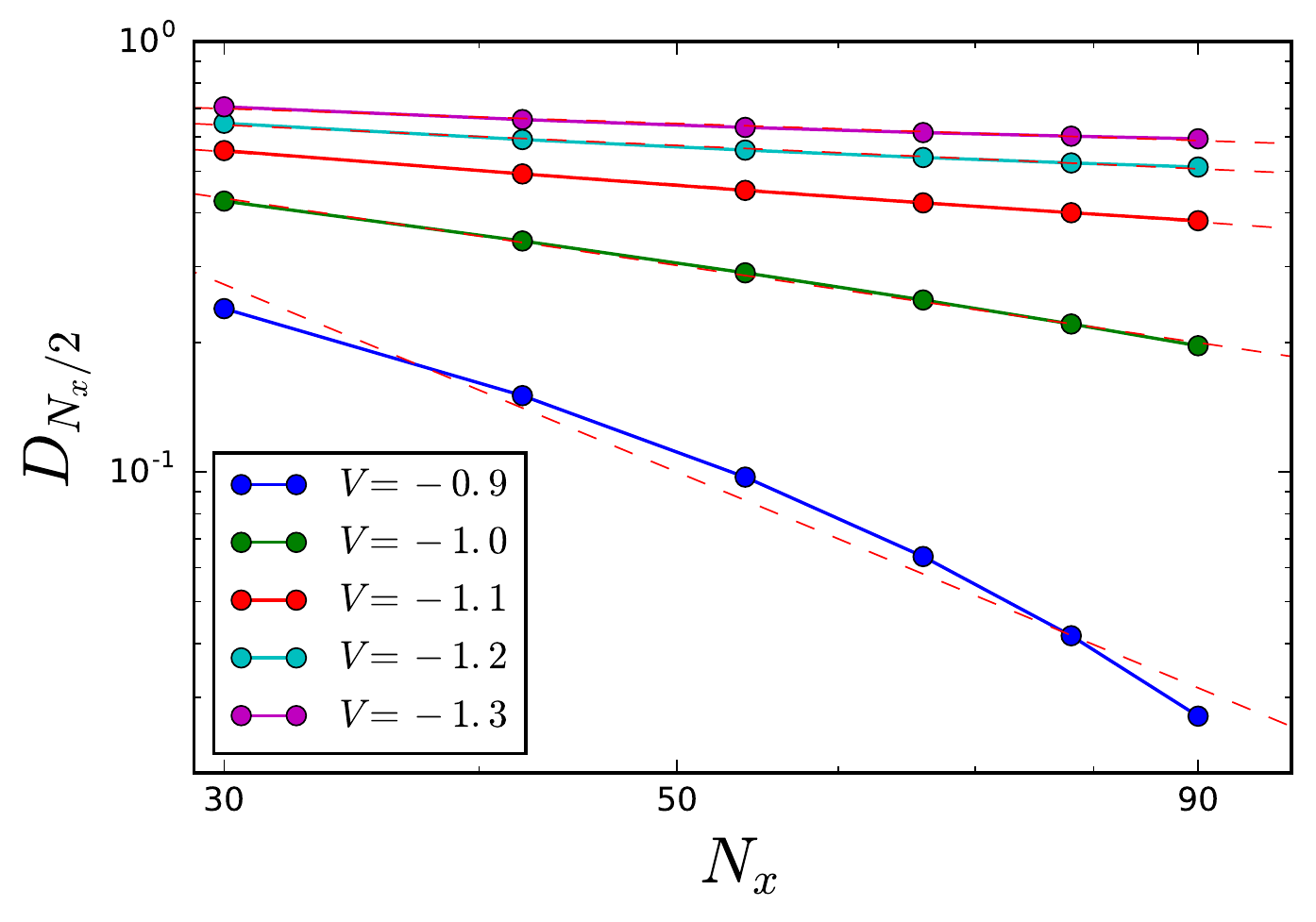}}

\subfloat[\label{fig:dimer_dvsV}]{\includegraphics[width=0.7\columnwidth]{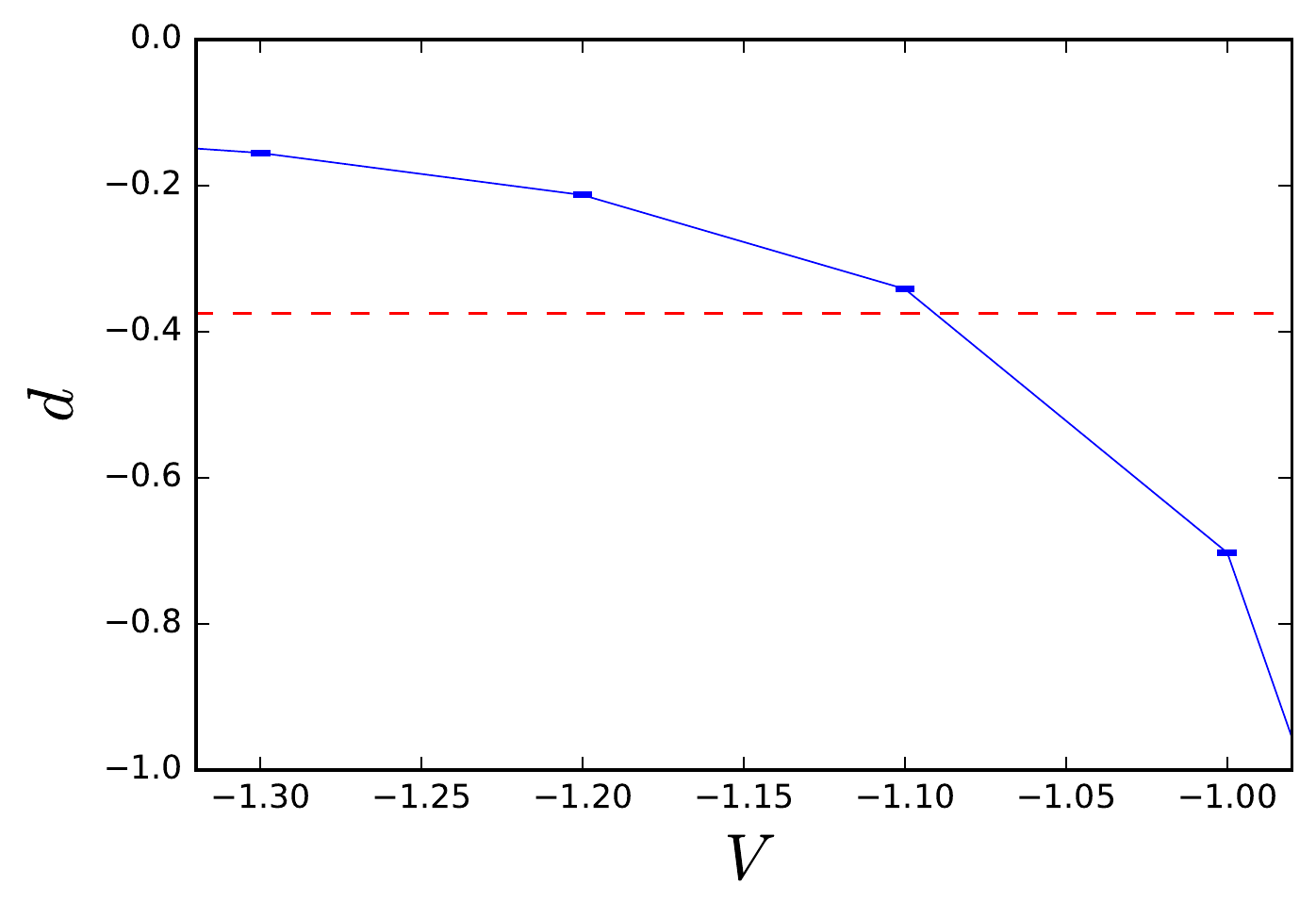}}

\caption{(a) The local dimerization at the central bond, as function of system
size, for $U=2$ and varying $V$. For each value of $V$, a fit to
a power law is plotted with a red dashed line. (b) The exponent, extracted
from the fit of the dimerization at the central bond as function of
system size to a power law, as function of $V$. The red dashed line
corresponds to $d=3/8$ - the exponent expected at the dimerization
transition.}

\end{figure}

As was mentioned in the main text, we find that large attractive pairing
interactions drive a transition into a dimerized phase. To obtain
the point of the phase transition into the dimerized phase, we calculate
the local dimerization $D_{i}=\left|\vec{S}_{i}\cdot\vec{S}_{i+1}-\vec{S}_{i-1}\cdot\vec{S}_{i}\right|$. 
At the dimerization transition, the dimerization
in the middle of the chain is expected to decay as a power law $D_{N_{x}/2}\sim N_{x}^{-d}$
with an exponent $d=3/8$. Performing finite size scaling, we fit
$D_{N_{x}/2}$ to a power law (see Fig. \ref{fig:dimer_finitesize}),
extracting the exponent $d$ as function of $V$. We then identify the
phase transition point as the value of $V$ for which the exponent
equals $3/8$ (see Fig. \ref{fig:dimer_dvsV}). 

\subsection{Phase transition between the trionic and gapless Haldane phase and phase separation\label{appx:TrionicPT}}

\begin{figure}
\centering\includegraphics[width=0.7\columnwidth]{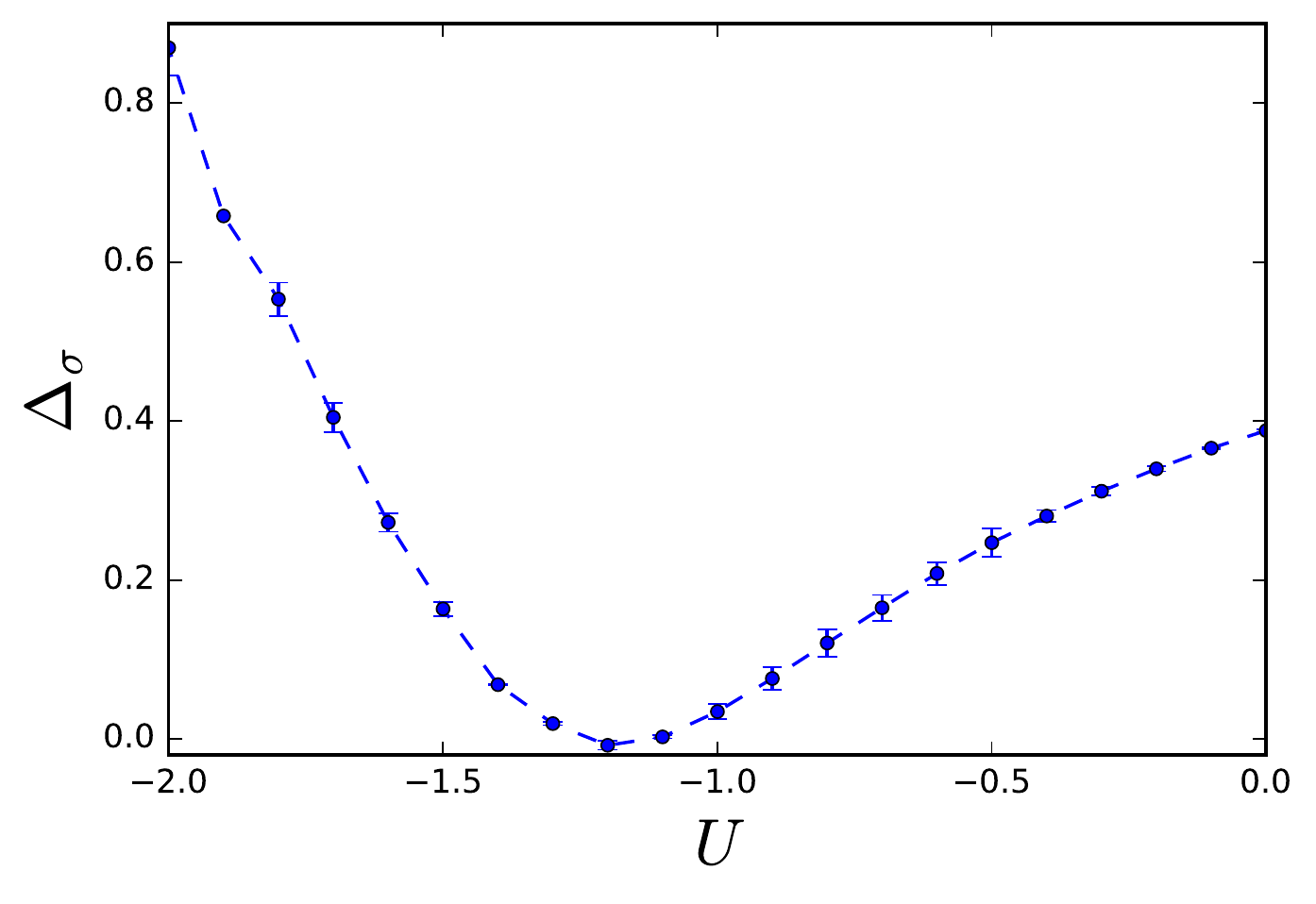}
\caption{ The spin gap, defined in Eq.~\eqref{eq:SpinGap}, for $V=-0.6$, as function of $U$. The
vanishingly small gap observed at the phase transition point numerically
is consistent with the weak coupling analysis.\label{fig:SpinGapNegU}}
\end{figure}

As was discussed in Sec. \ref{sec:Field-theory-analysis}, the phase
transition between the trionic phase and the gapless Haldane phase
is expected to be first order, with a finite spin gap at the transition.
In Fig. \ref{fig:SpinGapNegU}, we plot the spin gap as function of
$U$ for $V=-0.6$ and observe it approaching zero at the phase transition
point. We attribute this to the fact that, in the weak coupling limit,
the spin gap along the phase transition line is expected to be parametrically
smaller than deep in either phase, as was also mentioned in Sec. \ref{sec:Field-theory-analysis}.
To estimate the magnitude of the spin gap expected at the phase transition, one may use
 the one-loop RG equations and the results for the phase transition point obtained from DMRG. 
For instance, for $V=-0.6$, the phase transition is observed at $U_c \sim -1$, giving a spin gap $\Delta_{\sigma} \sim 0.5\cdot 10^{-3}$.
Going to larger system sizes, and larger bond dimension used in the
DMRG calculation, would perhaps allow one to resolve the finite size
of the gap at the transition.

\begin{figure}
\centering\includegraphics[width=0.7\columnwidth]{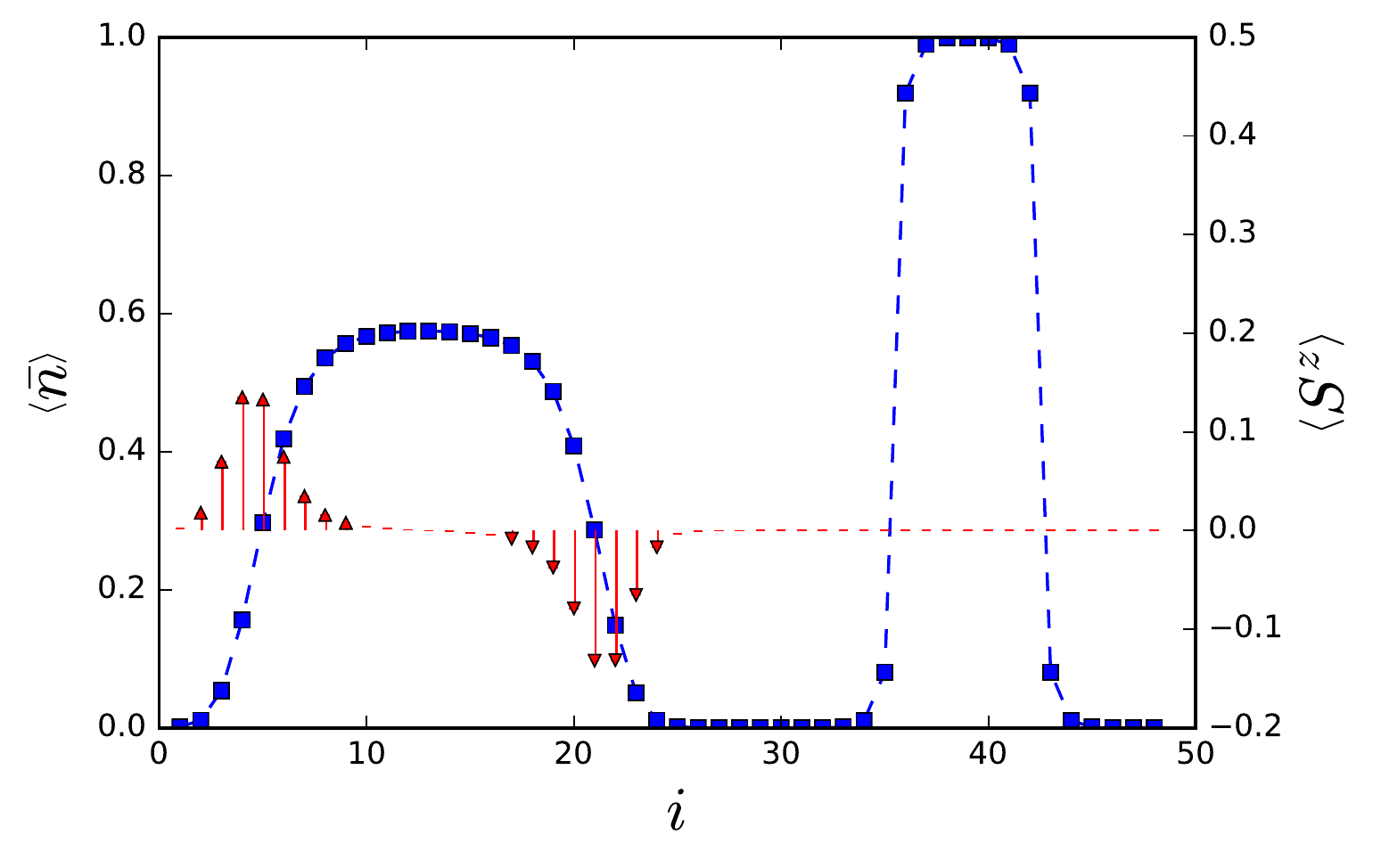}

\caption{Expectation value of the average density $\bar{n}=\left(n_{+1}+n_{0}+n_{-1}\right)/\left(3N_x\right)$
(blue squares) and the spin density (red arrows) in the phase separated
regime, for $V=-1.5$, $U=-1.6$. A coexistence between the topological
gapless Haldane phase, featuring localized spin-$1/2$ modes at its
boundaries, and the trivial trionic phase (with the trions bunching
together due to the large attractive pairing interactions), can be
observed.}
\label{fig:phase_sep}
\end{figure}

As was mentioned in the main text, for large $U,V<0$ the system tends to phase separate.
Close to the line of the phase transition between the gapless Haldane and the trionic phase the system phase
separates into regions in the gapless Haldane phase and a clustered
trions region, with localized spin-$1/2$ modes at the boundaries as can be seen in Fig. \ref{fig:phase_sep}.
This supports the statement that the phase transition between the gapless Haldane and
the trionic phases is indeed first order.

\subsection{Stability to $SO\left(3\right)$ symmetry breaking\label{appx:SIA}}

\begin{figure}
\centering\includegraphics[width=0.7\columnwidth]{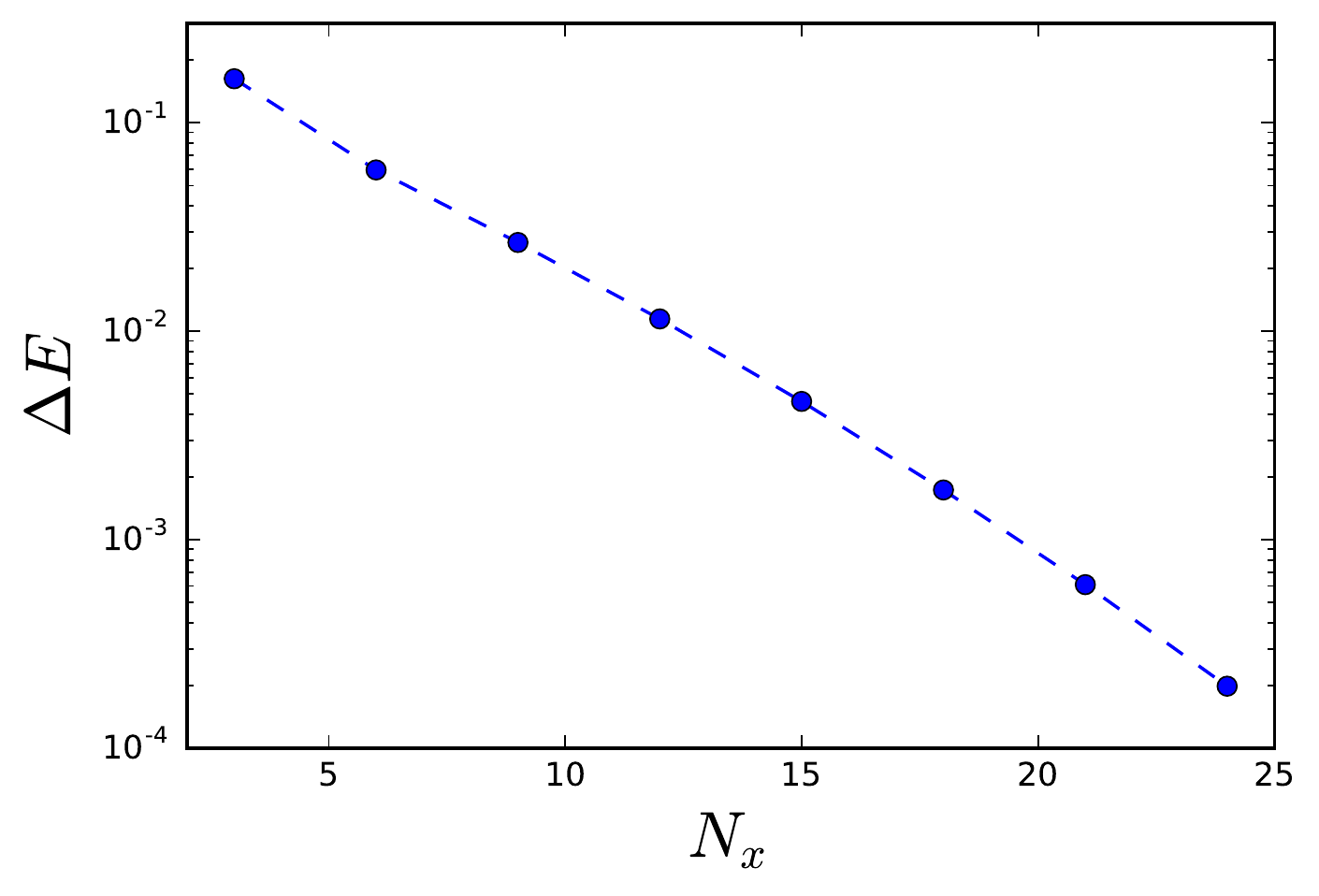}

\caption{The energy splitting between the states $\left|S_{L}^{z}=\frac{1}{2},S_{R}^{z}=\frac{1}{2}\right\rangle $
and $\left|S_{L}^{z}=\frac{1}{2},S_{R}^{z}=-\frac{1}{2}\right\rangle $,
in presence of a single ion anisotropy term $\sum_{i}D_{z}\left(S_{i}^{z}\right)^{2}$,
for $D_{z}=0.4$, as function of system size.\label{fig:SIA}}

\end{figure}

As discussed in the main text, 
we expect the topological phase to be stable to breaking of $SO\left(3\right)$ symmetry, 
as long as the $(\mathbb{Z}_2)^3$ symmetry, associated with the 
conservation of the fermionic parities of the three species, is preserved. 
To test this, we add a single ion anisotropy term $\sum_{i}D_{z}\left(S_{i}^{z}\right)^{2}$ to the model \eqref{HSO(N)}, 
and study the coupling between the end-modes.
To this end, we calculate the energy
splitting between the states $\left|S_{L}^{z}=\frac{1}{2},S_{R}^{z}=\frac{1}{2}\right\rangle $
and $\left|S_{L}^{z}=\frac{1}{2},S_{R}^{z}=-\frac{1}{2}\right\rangle $
as function of system size. 
(To induce the desired polarization at each end, we apply a small Zeeman field
at both ends of the chain.)
As can be seen in Fig. \ref{fig:SIA}, we find the energy
splitting, and hence the coupling between the end modes, to be exponential in system size.

\end{document}